\documentclass[lettersize,journal]{IEEEtran}
\usepackage{array}
\usepackage[caption=false,font=normalsize,labelfont=sf,textfont=sf]{subfig}
\usepackage{textcomp}
\usepackage{stfloats}
\usepackage{url}
\usepackage{verbatim}
\usepackage{cite}
\usepackage{tikz}  
\usepackage{verbatim}
\usetikzlibrary{arrows}
\usetikzlibrary{shapes.arrows}
\usetikzlibrary{backgrounds}
\usepackage{epstopdf}
\usepackage{epsfig}   
\usepackage{lscape} 
\usepackage{color}\definecolor{MyDarkBlue}{rgb}{0,0.08,0.45}\definecolor{yellow}{rgb}{0.99,0.99,0.70}\definecolor{white}{rgb}{1.0,1.0,1.0}\definecolor{black}{rgb}{0.00,0.00,0.00}

\pagecolor{white}
\usepackage{geometry}
\makeatletter
\def\ps@headings{%
\def\@oddhead{\mbox{}\scriptsize\rightmark \hfil \thepage}%
\def\@evenhead{\scriptsize\thepage \hfil \leftmark\mbox{}}%
\def\@oddfoot{}%
\def\@evenfoot{}}
\makeatother
\usepackage{graphicx}
\usepackage{algorithm}
\usepackage{algpseudocode}
\usepackage{enumerate} 
\usepackage{epsfig} 
\usepackage{CJK} 
\usepackage{amsmath} 
\usepackage[para]{threeparttable}
\usepackage{booktabs} 
\usepackage{amsmath} 
\usepackage{amsfonts} 
\usepackage{upgreek}    
\usepackage{bm} 
\usepackage{mathrsfs}   
\usepackage{multirow} 
\usepackage{siunitx} 
\usepackage{amssymb} 

\newcommand\ASTART{\bigskip\noindent\begin{minipage}[b]{0.5\linewidth}}
\newcommand\ACONTINUE{\end{page}\begin{minipage}[b]{0.5\linewidth}}
\newcommand\AENDSKIP{\end{minipage}\bigskip}
\newcommand\AEND{\end{minipage}}

\ifCLASSINFOpdf
\else
\fi
\begin{document}
\bibliographystyle{ieeetran}
\title{Blockage-Aware Robust Beamforming in RIS-Aided Mobile Millimeter Wave MIMO Systems}  
\author{Yan~Yang,~\IEEEmembership{Senior Member,~IEEE,}
          Shuping~Dang,~\IEEEmembership{Senior Member,~IEEE,}
        Miaowen~Wen,~\IEEEmembership{Senior Member,~IEEE,}
        Bo~Ai,~\IEEEmembership{Fellow,~IEEE,}
        and~Rose Qingyang Hu,~\IEEEmembership{Fellow,~IEEE}
\thanks{Part of this manuscript has been accepted for publication in IEEE ICC 2024 \cite{Yang24}. This work was funded by the National Nature Science Foundation of China under Grants 62271041 and 62221001. \textit{(Corresponding author: Yan Yang)}}
\thanks{Yan Yang and Bo Ai are with the School of Electronic and Information Engineering, Beijing Jiaotong University, Beijing 100044, China (email: yyang@bjtu.edu.cn, boai@bjtu.edu.cn). }
\thanks{Shuping Dang is with the School of Electrical, Electronic and Mechanical Engineering, University of Bristol, Bristol BS8 1UB, U.K. (e-mail: shuping.dang@bristol.ac.uk). }
\thanks{Miaowen Wen is with the School of Electronic and Information Engineering, South China University of Technology, Guangzhou 510640, China (e-mail: eemwwen@scut.edu.cn).}
\thanks{Rose Qingyang Hu is with the Department of Electrical and Computer Engineering, Utah State University, Logan, UT 84322, USA (e-mail: rose.hu@usu.edu).}}

\markboth{Journal of \LaTeX\ Class Files,~Vol.~14, No.~8, August~2021}%
{Shell \MakeLowercase{\textit{et al.}}: A Sample Article Using IEEEtran.cls for IEEE Journals}
%
\maketitle
\thispagestyle{empty} 
\begin{abstract}
Millimeter wave (mmWave) communications are sensitive to blockage over radio propagation paths. The emerging paradigm of reconfigurable intelligent surface (RIS) has the potential to overcome this issue by its ability to arbitrarily reflect the incident signals toward desired directions. This paper proposes a Neyman-Pearson (NP) criterion-based blockage-aware algorithm to improve communication resilience against blockage in mobile mmWave multiple input multiple output (MIMO) systems. By virtue of this pragmatic blockage-aware technique, we further propose an outage-constrained beamforming design for downlink mmWave MIMO transmission to achieve outage probability minimization and achievable rate maximization. To minimize the outage probability, a robust RIS beamformer with variant beamwidth is designed to combat uncertain channel state information (CSI). For the rate maximization problem, an accelerated projected gradient descent (PGD) algorithm is developed to solve the computational challenge of high-dimensional RIS phase-shift matrix (PSM) optimization. Particularly, we leverage a subspace constraint to reduce the scope of the projection operation and formulate a new Nesterov momentum acceleration scheme to speed up the convergence process of PGD. Extensive experiments confirm the effectiveness of the proposed blockage-aware approach, and the proposed accelerated PGD algorithm outperforms a number of representative baseline algorithms in terms of the achievable rate. 
\end{abstract}
\begin{IEEEkeywords}
Blockage detection, millimeter wave (mmWave) communications, reconfigurable intelligent surfaces (RIS), robust beamforming, projected gradient descent, momentum acceleration.
\end{IEEEkeywords}
%

\IEEEpeerreviewmaketitle
\section{Introduction}
\IEEEPARstart{A}{s} 
the front runner of future wireless techniques, millimeter wave (mmWave) communication has attracted considerable research attention since a large chunk of potentially usable bandwidth can be unleashed to support the ever-increasing data-intensive applications\cite{Rappaport19}. Meanwhile, signal transmissions over mmWave bands are susceptible to high path and penetration loss. By virtue of large-scale antenna arrays, path loss at mmWave bands can be compensated by highly directional beamforming at both transmitter and receiver, by which a very narrow beam pattern is used to provide a high directional antenna gain. At present, hybrid analog-digital (HAD) beamforming architecture is a favorable solution, which is capable of achieving almost the same performance as the full-digital beamforming scheme\cite{Heath16,Alkhateeb14-2}. However, the highly directional mmWave transmission is sensitive to the blockage over the radio propagation path. The presence of random blockage, as well as self-blockage, e.g., human-body blockage, normally incurs a sharp degradation of signal strength, resulting in a high outage probability\cite{Raghavan23}. Hence, mmWave communications are subject to a stringent limitation imposed by the blockages and are likely to only operate in line-of-sight (LoS) scenarios. In mobile mmWave communication networks, it has been proven that intermittent connectivity is rather detrimental\cite{Jain19}.
\\
\indent
To overcome such a challenge, the recent emergence of reconfigurable intelligent surfaces (RISs), a.k.a. intelligent reflecting surfaces (IRSs), provides a potential solution by utilizing their unnatural reflection property\cite{ Bjorson20, Wu19,Zhao21}. Specifically, the surface is composed of a large number of low-cost and nearly passive elements, each capable of electronically controlling the phase of incident electromagnetic waves (EMs). By actively manipulating phase shifts (PSs), the reflection behavior of EMs can be purposefully regulated. Therefore, RIS can reflect incident signals toward desired directions, thereby endowing radio channels with programmability\cite{Chen20, Chen16, Renzo20, Basar19}. Unlike traditional amplify-and-forward (AF) relays and coordinated multi-point (CoMP) transmission, RIS is almost passive. As such, even though massive RISs are densely deployed, sophisticated interference management could be exempted. Such a deployment has the potential to give rise to a dramatic improvement in the overall network's signal-to-interference-plus-noise ratio (SINR) performance. 
\\
\indent
For highly directional mmWave communications, the validity of an RIS substantially depends on whether the blockages over the radio propagation path can be effectively detected. More precisely, the receiver should be sufficiently aware of the blockage status of a link, i.e., whether the blockage is present and under what condition. In this context, when a complete link obstruction arises, a base station (BS) can electronically steer the beam toward a desired RIS, thereby providing an indirect link from the RIS to the corresponding mobile station (MS). Prior works in the literature have focused on blockage prediction, robust beamforming, and outage minimization to overcome such challenges. With the aid of CoMP transmission, a stochastic learning approach was proposed to capture crucial blockage patterns, and a robust beamforming design was established to combat uncertain path blockages, e.g.,\cite{ Kumar21,Jiao22}. Similarly, a learning-based robust beamforming design was proposed for RIS-aided mmWave systems in the presence of random blockages \cite{Zhou21}. 
\\
\indent
In this new paradigm, because the RIS comprises only nearly passive elements and has limited signal processing capability, the acquisition of channel state information (CSI) is much more challenging compared with the case without an RIS\cite{Zhou20, Wang20,Pan23, Shi22}. On the one hand, RIS-aided communications involve the cascaded BS-to-RIS and RIS-to-MS channels, where the number of channel coefficients is proportional to the number of elements. Therefore, channel estimation becomes a complicated problem due to the involvement of thousands of interrelated variables\cite{Zhou20,Pan23}. In most existing works, the perfect CSI is assumed to be obtained for optimally designing the precoding vectors at the BS and the RIS phase-shift matrix (PSM). Nevertheless, such an idealistic assumption is practically infeasible. Many recent research efforts have attempted to develop new approaches for channel estimation, reflection optimization, and robust beamforming. Notably, the authors in \cite{Perovic21} has considered these critical issues, as well as pursue rate maximization with lower complexity. On the other hand, in \cite{Jiao22}, the authors verified that by exploiting the sparse mmWave channel structure, the channel coefficients of reflective channels can be derived based on the estimation of angle-of-departure (AoD), angle-of-arrival (AoA), and corresponding channel gains. At present, despite these advances, relatively scarce studies have considered the uncertainties of mmWave radio channel and random blockages, as well as the impact of imperfect CSI on the performance of RIS-aided mmWave communication systems. It is worth noting that the presence of random blockages may lead to more unpredictable channel uncertainties, exacerbating the impact of the imperfect CSI.
\\
\indent
Motivated by this, we focus on tackling the impact of the blockages on RIS-aided mmWave mobile communication systems. Specifically, we develop a pragmatic blockage-aware algorithm to substantially reap the benefits of the RIS, such that when the LoS link is blocked, the BS can immediately steer the dominant beam towards the RIS and create an alternative reflection link. In contrast, if the blockage status cannot be proactively identified, the role of RIS would be futile, since effective blockage detection is a fundamental prerequisite for RIS reflection to take effect. Furthermore, we propose a robust beamforming design for downlink mmWave MIMO transmission to achieve outage probability minimization and achievable rate maximization. The main contributions of this work are summarized as follows: 
\begin{itemize}
\item From the perspective of the statistical decision theory, we transform the blockage detection problem into a Neyman-Pearson (NP) criterion-based binary hypothesis testing problem. In this way, we develop an NP criterion-based blockage detector and a trusted NP decision rule. The goal is to maximize the probability of correct detection and exclude the worst-case system outage. Experiments validate our theoretical findings and show that without a priori knowledge of randomly distributed blockers, the proposed blockage-aware approach can accurately distinguish whether a channel is LoS or non-line-of-sight (NLoS), giving the opportunity to exclude potential outages preemptively. To the best of our knowledge, the proposed NP decision rule is the first trustworthy solution to the random blockage detection problem in mmWave communication systems. 
\item Additionally, by taking advantage of angle reciprocity, we propose a mechanism of link availability indication and formulate an easy-to-handle geometric transformation approach, which can be integrated into the transceiver and helps to mitigate the impact of the uncertain CSI introduced by cascaded BS-RIS-MS channels.
\item By virtue of this blockage-aware technique, we further propose an outage-constrained robust beamforming scheme. Specifically, a beamwidth-variant RIS beamformer is designed to combat the uncertainty of cascaded BS-RIS-MS channels. Compared to the constant-beamwidth strategy, our key observation is that even though the estimates of the CSI and/or PSs are relatively coarse, the RIS can provide more resilient connectivity for blocked users. The simulation results show that even though a broadened beam may temporarily reduce the achievable rate, the sum rate is higher than those of the baseline algorithms and is close to the optimum achievable rate due to significant suppression of the outage probability and beam-training overhead.
\item To solve the computational challenge of high-dimensional PSM optimization for large-scale RIS reflection arrays, we propose an accelerated projected gradient descent (PGD) method with a best-fit subspace constraint. Essentially, this accelerated version of PGD employs a simplex Euclidean projection onto a feasible convex set for the constrained PSM optimization problem. We further leverage a subspace constraint to reduce the scope of the projection operation, making the aforementioned intractable problem solvable. To speed up the convergence process, we propose a new Nesterov momentum acceleration scheme. We demonstrate the superiority of this scheme that can avoid local minima by preventing the PGD algorithm from getting stuck at the vicinity saddle points. Unlike the traditional alternating optimization (AO) algorithm, which only updates a single variable in each iteration, the accelerated PGD algorithm proposed in this paper can simultaneously update all optimization variables, resulting in faster convergence and a higher achievable rate. 
\end{itemize}

The remainder of the paper is organized as follows. In Section II, we introduce the system and channel models and formulate the problems of the outage probability minimization and achievable rate optimization. In Section III, we provide a general statistical decision-making framework to enable the awareness of blockage and devise a mechanism for link availability indication. In Section IV, a robust beamformer is developed to mitigate the impact of the channel uncertainty, and a PGD-based solution is derived to maximize the achievable rate. Simulation results are shown and discussed in Section V, and the paper draws conclusions in Section VI. 
\\
\underline {\textit{Notations}}: Throughout the paper, bold upper and lower case letters denote matrices and vectors, respectively; normal lower case letters are used for scalars; Superscripts $(\cdot)^*$, $(\cdot)^T$ and $(\cdot)^H$ stand for the complex conjugation, transpose, and Hermitian transpose, respectively. We use $\mathbb C$ to denote the field of complex numbers and $\mathbb C^{m \times n}$ to denote an $m$ by $n$ dimensional complex space. Further, we use ${\text{Tr}}(\cdot)$, ${\text{log}}(\cdot)$, and $\vert \cdot \vert$ to denote the trace, the natural logarithm, and the determinant of a matrix or the absolute value depending on the context, respectively. ${\mathcal {CN}}(0;\sigma^2)$ represents the zero-mean complex Gaussian distribution with variance $\sigma^2$. Finally, $\nabla_{\bm\theta}\mathcal R(\bm\theta)$ denotes the partial gradient of $\mathcal R(\cdot)$ with respect to parameter $\bm\theta$. 
\section{System Model and Problem Formualtion}
\subsection{Signal Transmission and Reception}
\begin{figure*}[!t]
\centering
\includegraphics[width=16.5cm]{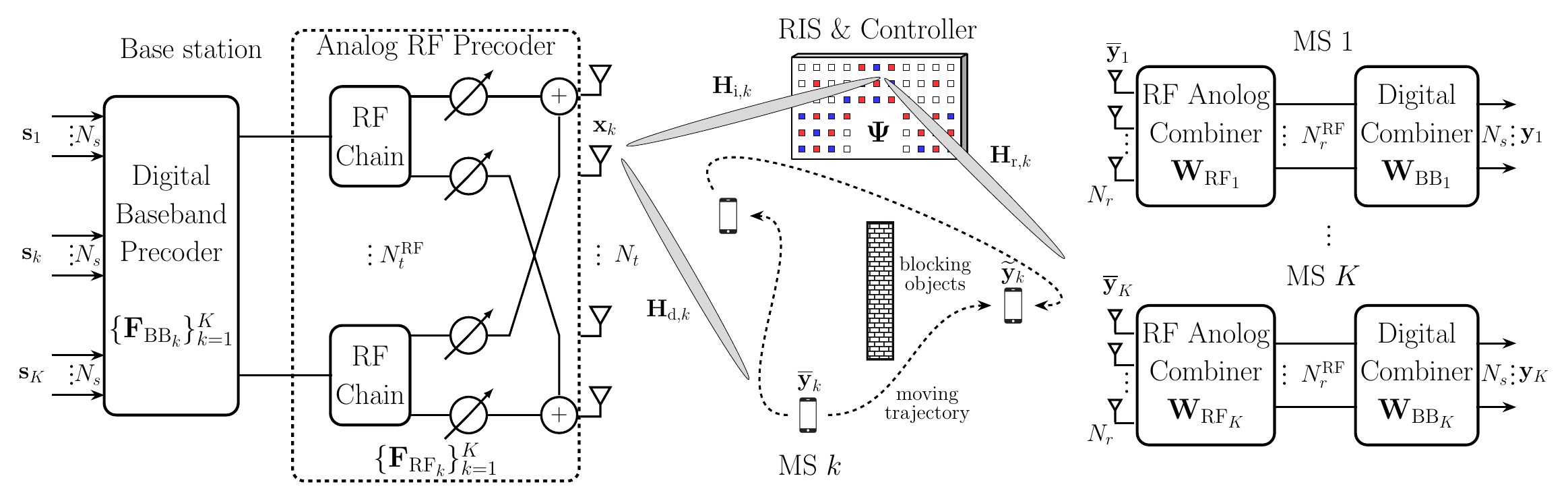}
\caption{RIS-aided mmWave MIMO communication system with HAD beamforming: A downlink transmission subject to blockages.}
\label{figure}
\end{figure*}
We consider an RIS-aided mmWave MIMO communication system in the downlink, as illustrated in Fig. 1, where the RIS, with $\mathcal N$ reflecting elements arranged in an ${\mathcal I} \times {\mathcal J}$ uniform rectangular array (URA) layout, is utilized to provide a substitutable link in the presence of random link blockages. A BS with a two-stage HAD beamforming architecture uses a uniform linear array (ULA) of $N_t$ transmit antennas and $N_t^{\text{RF}}$ RF chains to serve $K$ MSs. Unless otherwise specified, we assume that each MS is equipped with $N_r$-antenna ULA and $N_r^{\text{RF}}$ RF chains, and these $N_{t,k}^{\text{RF}}$ RF chains can be configured for a specific MS to transmit varying $N_s$ data streams given $N_s\le N_t^{\text{RF}}$. During downlink transmission, the data stream ${\textbf s}=\{{\textbf s}_k\}_{k=1}^{K}, where {\textbf s}_k\in {\mathbb C}^{N_s\times 1}$, is processed by a set of digital baseband precoders $\{{\textbf F}_{\text{BB}_k}\}_{k=1}^K, where {\textbf F}_{\text{BB}_k}\in \mathbb C^{N_{t,k}^{\text{RF}} \times N_s}$, and followed by a set of analog RF precoders $\{{\textbf F}_{\text{RF}_k}\}_{k=1}^K, where {\textbf F}_{\text{RF}_k}\in \mathbb C^{N_{t,k} \times N_s}$. Likewise, a hybrid combiner is deployed at the receiver by the concatenation of an RF combiner ${{\textbf W}_{{\text {RF}}_k}}\in \mathbb C^{N_{r,k} \times N_r^{\text{RF}}}$ and a low-dimensional digital baseband combiner ${\textbf W}_{{\text {BB}}_k}\in \mathbb C^{N_r^{\text{RF}} \times N_s}$ to sequentially process the digitally precoded streams. 
\\
\indent
Under such a setting, the RIS associated with an RIS controller is installed at a known location\footnote{The passive beamforming at the RIS is generally manipulated by the BS. As such, the BS computes the PSs at the RIS, and the corresponding results can be transferred to the RIS controller over dedicated control channels \cite{Pan23}.}. We define $\mathcal K=\{1,2,...,K\}$ to be the set of $K$ active MSs. For an arbitrary MS $k\in \mathcal K$, when the direct link is blocked, we suppose that $\mathcal N_k$ reflecting element of the RIS can be partitioned into multiple rectangular subarrays of size ${{\mathcal I_k} \times {\mathcal J_k}}$, given $\sum_{k=1}^K \mathcal N_k\le \mathcal N$, ${{\mathcal I}_k}\le {\mathcal I}$ and ${{\mathcal J}_k}\le {\mathcal J}$. ${\bm\Psi}_k({\bm\theta})= {\textrm{diag}}({\bm\theta}) \in \mathbb C^{{\mathcal N}_k\times {\mathcal N}k}$ is the PSM associated to MS $k$, where ${\bm\theta}=\left [e^{j\theta_{1,k}},\dots,e^{j\theta_{{\imath\jmath},k}},\dots\right ]^T \in \mathbb C^{{{\mathcal N_k}}\times1}$, $\{\theta_{\imath\jmath,k}\}_{\imath\jmath=1}^{\mathcal N_k}$ represents a set of PSs induced by these $\mathcal N_k$ RIS elements, and $\theta_{\imath\jmath,k}$ is the PS of the $(\imath,\jmath)$th entry of the URA, given $1\le\imath \le\mathcal I_k$ and $1\le\jmath \le\mathcal J_k$. Due to the sparse nature of mmWave channels, we can make a dichotomous assumption that the mmWave channel is either unblocked or blocked. This strong duality suggests that $\mathcal K$ could be partitioned into two disjoint subsets, denoted by $\mathcal K^{\text{NLoS}}$ and $\mathcal K^{\text{LoS}}$, corresponding to those MSs under the NLoS and the LoS conditions, respectively.
\\
\indent
In the downlink, the transmitted signal to the $k$th MS in a generic HAD mmWave system is given by
\begin{equation}
\label{eqn_BF_x_k}
{\textbf x}_k ={\textbf F}_k{{\textbf s}_k},\quad k\in\mathcal K,
\end{equation}
where ${\textbf F}_k ={\textbf F}_{\text{RF}_k} {\textbf F}_{\text{BB}_k}\in \mathbb C^{N_r \times N_{t,k}}$ is the hybrid precoder and is in nature different for each individual MS; ${{\textbf s}_k}\in \mathbb C^{N_s \times 1}$ is a normalized transmit signal vector, i.e., $\mathbb E\{\vert\vert{{\textbf s}_k} \vert\vert^2\}=1$. 
\\
\indent
Adopting the Rician fading channel model, both the direct path and the reflection path are considered, such that the channel matrix between the BS and the $k$th MS can be described as 
\begin{subequations}
\begin{equation}
\label{eqn_BF_H_k}
\overline{\textbf H}_k({\bm\theta}) =\varepsilon_k\sqrt{\frac{\kappa\overline\varrho_{{\text d},k}}{\kappa+1}}{{{\textbf H}}_{{\text{d}},k}}+\sqrt{\frac{\overline\varrho_{\text{c},k}}{\kappa+1}}{{\textbf H}}_{\text{c},k}({\bm\theta}), \;k\in{\mathcal K}^{\text{LoS}}
\end{equation}
\begin{equation}
\label{eqn_BF_H_k}
\widetilde{\textbf H}_k({\bm\theta}) =\varepsilon_k\sqrt{\frac{\widetilde\varrho_{{\text d},k}}{\kappa+1}}{{\textbf H}_{{\text{d}},k}}+\sqrt{\frac{\kappa\widetilde\varrho_{\text {c},k}}{\kappa+1}}{{\textbf H}}_{{\text c},k}({\bm\theta}),\;k\in{\mathcal K}^{\text{NLoS}}
\end{equation}
\end{subequations}
where ${{\textbf H}_{{\text{d}},k}}\in \mathbb C^{N_r\times N_{t,k}}$ represents the direct BS-MS channel, and ${\textbf H}_{\text{c},k}({\bm\theta})={{{\textbf H}_{\text{r},k}^H}{{\bm {\Psi}}_k}}(\bm\theta){\textbf H}_{\text{i},k}\in \mathbb C^{N_r\times N_{t,k}}$ represents the cascaded BS-RIS-MS channel, both associated with MS $k$. Here ${{\textbf H}_{\text{r},k}}\in \mathbb C^{\mathcal N_k \times N_t}$ and ${\textbf H}_{\text{i},k}\in \mathbb C^{\mathcal N_r \times {\mathcal N}_k}$ are the matrices corresponding to the indirect channels between the BS and the RIS (BS-RIS) and the RIS and MS (RIS-MS), respectively. In addition, $\kappa\in [0,\infty)$ is the non-centrality parameter of the Rician distribution; $\varepsilon_k\in[0,1]$ is a normalized random variable that represents the link loss rate caused by a blocker, and $\varepsilon_k=0$ corresponds to the worst-case channel condition, that is, the case where the $k$th MS is fully blocked. $\varrho_{\text d,k}$ and $\varrho_{\text r,k}$ are the fading coefficients of the direct and indirect links, respectively. To simplify the notations and facilitate the following analysis, we denote $\upsilon_0=\sqrt{\frac{\kappa\varrho_{\textup d,k}}{\kappa+1}}$ and $\upsilon_1=\sqrt{\frac{\varrho_{\text {c},k}}{\kappa+1}}$ by the two random variables accounting for the effects of both the path loss and the Ricianfactor factor in sparse scattering environments.
\\
\indent
Accordingly, the $N_r\times 1$ received signal vector at the antenna array of the $k$th MS can be expressed as
\begin{subequations}
\begin{equation}
\label{eqn_Revsignal_MSa}
\begin{aligned}
{\overline{\textbf y}}_k =  {\underbrace{\varepsilon_k\overline\upsilon_0{{\textbf W}^H _k}{{\textbf H}_{{\text{d}},k}}{{\textbf x}_k}}_{\text{Dominant signal}: \,{\overline{\textbf y}}_{{\text d},k}}}+{\underbrace{\overline\upsilon_1{{\textbf W}^H _k}{{{\textbf H}_{\text{r},k}^H}{{\bm {\Psi}}_k}(\bm\theta){{\textbf H}}_{\text{i},k}{{\textbf x}_k}}}_{\text{Reflected auxiliary signal}:\, {\overline{\textbf y}}_{{\text r},k}}+ {\textbf n}_k},
\end{aligned}
\end{equation}
\begin{equation}
\label{eqn_Revsignal_MSb}
\begin{aligned}
{\widetilde{\textbf y}}_k =  {\underbrace{{{\varepsilon_k\widetilde\upsilon_0}}{{{\textbf W}^H _k}{{\textbf H}}_{{\text{d}},k}}{{\textbf x}_k}}_{\text{Auxiliary signal}:\,{\widetilde{\textbf y}}_{{\text d},k}}}+{\underbrace{\widetilde\upsilon_1{{{\textbf W}^H _k}{{\textbf H}_{\text{r},k}^H}{{\bm {\Psi}}_k}(\bm\theta){{\textbf H}}_{\text{i},k}{{\textbf x}_k}}}_{\text{Reflected dominant signal}:\, {\widetilde{\textbf y}}_{{\text r},k}}+ {\textbf n}_k},
\end{aligned}
\end{equation}
\end{subequations}
where ${{\textbf W}^H _k}={{\textbf W}^H _{{\text {BB}}_k}}{{\textbf W} _{{\text {RF}}_k}}$; ${\textbf n}_k \sim \mathcal {CN}\left\{0,{\sigma}_k^2 \textbf I\right\}$ is the circularly symmetric complex additive white Gaussian noise (AWGN) at the $k$th MS; ${\sigma}_k^2$ represents the noise variance associated with the $k$th MS; ${\textbf I}\in \mathbb C^{N_r\times N_r}$ is the identity matrix. For ease of presentation, we also denote ${\textbf x}_{\text d,k}$ and ${\textbf x}_{\text r,k}$ as the direct and indirect signal impinging on the antenna array, respectively.
For convenience, we denote ${\textbf U}_k={{\textbf F}}^H_k{\textbf F}_{k}$ as the transmit beamforming gain, ${\textbf V}_k={{\textbf W}_k^H}{{\textbf W}_k}$ as the receive beamforming gain, and ${\textbf Z}_k(\bm\theta)={{{\textbf H}_{\text{r},k}^H}{{\bm {\Psi}}_k}(\bm\theta){\textbf H}_{\text{i},k}}$. The total beamforming gain is thus given by ${\textbf G}={\textbf V}_k{\textbf U}_k$. The signal-to-interference-plus-noise ratio (SINR) on the $k$th MS can now be defined in a closed form as
\begin{subequations}
\begin{equation}
\label{eqn_SINR_ka}
\begin{aligned}
\overline{\Gamma}_k(\bm\theta,{\textbf G})=&
\frac{\varepsilon_k^2\overline\upsilon_0^2}{J_0}\left\vert{\textbf H}_{\text d,k}{\textbf G}{\textbf H}_{\text d,k}^H\right\vert+\frac{\overline\upsilon_1^2}{J_0}\left\vert{\textbf Z}_k(\bm\theta){\textbf G}{\textbf Z}_k^H(\bm\theta)\right\vert,\;\;
\end{aligned}
\end{equation}
\begin{equation}
\label{eqn_SINR_kb}
\begin{aligned}
\widetilde{\Gamma}_k(\bm\theta,{\textbf G})=&
\frac{\varepsilon_k^2\widetilde\upsilon_0^2}{J_0}\left\vert{\textbf H}_{\text d,k}{\textbf G}{\textbf H}_{\text d,k}^H\right\vert+\frac{\widetilde\upsilon_1^2}{J_0}\left\vert{\textbf Z}_k(\bm\theta){\textbf G}{\textbf Z}_k^H(\bm\theta)\right\vert,
\end{aligned}
\end{equation}
\end{subequations}
%
where $J_0=N_{t,k}\sigma_k^2+\sum_{j\ne k}g_j$ is the total signal power received from all the other MSs, and $\sigma_k^2$ is a measure of the noise at the $k$th MS. Under the assumption that the perfect CSI is available at the transceiver, the achievable rate of the $k$th user, denoted by $\mathcal R_k$ and measured in bit/s/Hz, can be calculated as 
\begin{subequations}
\begin{equation}
\label{eqn_R_ka}         
\begin{aligned}
\overline{\mathcal R}_k(\bm\theta,\textbf G)&={\log_2}\det\left({\textbf I}_k+\frac{1}{J_0}\overline{\textbf H}_k({\bm\theta}){\textbf G}{\overline{\textbf H}}_k^H({\bm\theta})\right),
\end{aligned}
\end{equation}
\begin{equation}
\label{eqn_R_kb}         
\begin{aligned}
\widetilde{\mathcal R}_k(\bm\theta,\textbf G)&={\log_2}\det\left({\textbf I}_k+\frac{1}{J_0}\widetilde{\textbf H}_k({\bm\theta}){\textbf G}{\widetilde{\textbf H}}_k^H({\bm\theta}) \right).
\end{aligned}
\end{equation}
\end{subequations}
In such a case, the downlink sum rate can be determined as
\begin{equation}
\begin{aligned}
\label{eqn_capacity_sys}
 {\mathcal R}_{\text{sys}} ={\sum\limits}_{k\in \mathcal {K}^{\text{LoS}}} \overline{\mathcal R}_k({\bm\theta},{\textbf G})+{\sum\limits}_{k\in \mathcal {K}^{\text{NLoS}}} \widetilde{\mathcal R}_k(\bm\theta,\textbf G).\end{aligned}
\end{equation}
\subsection{Channel Model}
Due to the inherent sparsity of mmWave channels, we adopt a geometric channel model as proposed in\cite{Akdeniz14} to describe the characteristics of path clusters, i.e., taps, in the angular domain between transceiver pairs. At the $k$th MS, suppose that the number of effective channel clusters for the BS-MS links, the BS-RIS links, and the RIS-MS links are $L_{\text{d},k}$, $L_{\text{r},k}$, and $L_{\text{i},k}$, respectively. With this clustered channel model, the channel matrices in (3) can be written as
\begin{equation}
\begin{aligned}
\label{eqn_nb_channelI}
{\textbf H}_{\text{d},k}={\frac{1}{\sqrt{L_{\text{d},k}}}} \sum \limits_{\ell=1}^{{ L}_{\text{d},k}}\textsl{g}^\ell_{\text{d},{k}}{\textbf a}_{{ \text{MS}}}\left(\psi^{\text{AoA}_\ell}_{{\text{d},{k}}}\right){\textbf a}_{{ \text{BS}}}\left(\phi^{{\text{AoD}_\ell}}_{{\text{d},{k}}}\right)^H,
\end{aligned}
\end{equation}
\begin{equation}
\begin{aligned}
\label{eqn_nb_channelI}
{\textbf H}_{\text{r},k}={\frac{1}{\sqrt{L_{\text{r},k}}}} \sum \limits_{\ell=1}^{L_{\text{r},k}}\textsl{g}^{\ell}_{\text{r},{k}}{\textbf a}_{{ \text{MS}}}\left(\psi^{\text{AoA}_\ell}_{{\text{r},{k}}}\right){\textbf a}_{{ \text{RIS}}}\left(\phi^{\text{AoD}_\ell}_{{\text{r},{k}}},\varphi^{\text{AoD}_\ell}_{{\text{r},{k}}}\right)^H,
\end{aligned}
\end{equation}
\begin{equation}
\begin{aligned}
\label{eqn_nb_channelI}
{\textbf H}_{{\text{i},k}}={\frac{1}{\sqrt{L_{\text{i},k}}}} \sum \limits_{\ell=1}^{L_{\text{i},k}}\textsl{g}^\ell_{{\text{i},k}}{\textbf a}_{{ \text{RIS}}}\left(\phi^{\text{AoA}_\ell}_{{\text{i},k}},\varphi^{\text{AoA}_\ell}_{{\text{i},k}}\right){\textbf a}_{{ \text{BS}}}\left(\phi^{\text{AoD}_\ell}_{{\text{d},k}}\right)^H,
\end{aligned}
\end{equation}
where $\textsl{g}^\ell_{\text{d},{k}}$, $\textsl{g}^{\ell}_{\text{r},{k}}$ and $\textsl{g}^\ell_{{\text{i},k}}$ are the complex gains of the $\ell$th tap of the corresponding links, respectively; $\left\{\psi^{\text{AoA}_\ell}_{{\text{d},{k}}},\phi^{{\text{AoD}_\ell}}_{{\text{d},{k}}}\right\}$, $\left\{\psi^{\text{AoA}_\ell}_{{\text{r},{k}}},\left\{\phi^{\text{AoD}_\ell}_{{\text{r},{k}}},\varphi^{\text{AoD}_\ell}_{{\text{r},{k}}}\right\}\right\}$ and $\left\{\left\{\phi^{\text{AoA}_\ell}_{{\text{i},k}},\varphi^{\text{AoA}_\ell}_{{\text{i},k}}\right\},\phi^{\text{AoD}_\ell}_{{\text{d},k}}\right\}$ are the azimuth (horizontal) AoA and AoD pairs of the $\ell$th tap for the BS-MS links, the RIS-MS links, and the BS-RIS links, respectively. Moreover, ${\textbf a}_{{ \text{BS}}}(\phi)$, ${\textbf a}_{{ \text{MS}}}(\phi)$ and ${\textbf a}_{{ \text{RIS}}}(\phi,\varphi)$ represent the transmit (BS), receive (MS) and reflect (RIS) array responses, i.e., steering vectors, respectively, given by
\begin{equation}
\begin{aligned}
\label{eqn_ris_channelI}
&{\textbf{a}}_{\text{BS}}({\phi}) = \frac{1}{{\sqrt {{N_{t,k}}} }}{\left[ {1,{e^{j{{\cos}}({{\phi}})}}},\dots,{e^{j({N_{t,k}}-1){{\cos}}({{\phi}})}}\right]^T,}
\\ 
&{\textbf{a}}_{\text{MS}}({\psi}) = \frac{1}{{\sqrt {{N_{r}}} }}\left[ {1,{e^{j{{\cos}}({\psi})}},\dots,{e^{j({N_{r}}-1){{\cos}}({\psi})}}} \right]^T,
\\ 
&{\textbf{a}}_{\text{RIS}}({\phi},\varphi) = \frac{1}{{\sqrt {{\mathcal N_{k}}} }}\left[1,{e^{j\xi_1(\phi,\varphi)}},\dots,{e^{j\xi_{\imath\jmath}(\phi,\varphi)}} \right]^T,
\end{aligned}
\end{equation}
where ${\xi}_{\imath\jmath}(\phi,\varphi)=(\imath-1){{\sin}}\,{\varphi}{{\sin}}{\phi}+(\jmath-1){{\cos}}{\phi}$ is the phase of an incoming plane wave at the $\imath\jmath$th RIS element; $\lambda$ is the wavelength, and $d$ is the inter-element distance of the RIS. For simplicity, we adopt a uniformly normalized geometric array factor to describe the layout of the radiating elements of BS, RIS, and MS.
\\
\indent
In general, the CSI acquisition of the cascaded channel is challenging due to the passive nature of RIS. However, from the perspective of the angular domain, the uncertainty of $\widetilde{\textbf H}_{\text{r},k}$ is significantly greater than that of $\widetilde{\textbf H}_{\text{d},k}$, and the channel estimation errors of the cascaded BS-RIS-MS links are mainly determined by the reflection links. A important reason is that the BS-RIS link is actually a fixed point-to-point cluster channel, which means that $\widetilde{\textbf H}_{\text{i},k}$ could be completely known in the angular domain. 
\\
\indent
Hence, the uncertainty of CSI in RIS-aided mmWave communication systems can be simplified to be solely described by the estimated error of the RIS-MS channel. As a result, estimated $\hat{\widetilde{\textbf H}}_{\text r,k}$ can be calculated as \cite{Rusek13}
\begin{equation}
\label{eqn_SE_EE_par} 
{\widetilde{\textbf H}}_{\text r,k}(\hat{\bm\theta})={{\varsigma}_k}{{\widetilde{\textbf H}}}_{\text r,k}({\bm\theta})+\sqrt{1-\varsigma_k^2} \textbf E, ~\forall \,k\in \mathcal K^{\text{NLoS}},
\end{equation}
where $\varsigma_k \in [0,1]$ represents the reliability of the estimate, and $\textbf E$ is an error matrix with i.i.d. $\mathcal {CN}(0,1)$ distributed entries.
\\
\indent
In this way, the uncertainty of the cascaded BS-IRS-MS channel can ultimately be reflected as inaccurate estimates of PSs. The mean squared error (MSE) of the channel uncertainty associated with the inaccurate estimates of PSs can be written as 
\begin{equation}
\label{eqn_SE_EE_par} 
\text {MSE}_{{\bm\theta}}= {\mathbb E}\left\{\left\vert\left\vert{\bm\theta}-{\hat{\bm\theta}}\right\vert\right\vert^2 \right\},~\forall \, k\in \mathcal K^{\text{NLoS}}.
\end{equation}
\subsection{Problem Formulation}
For potentially blocked user $k$, our goal here is to maximize the achievable rate by jointly adapting transmit beamformer ${\textbf U}_k$, receive beamformer ${\textbf V}_k$, and PSM $\bm\Psi_k(\bm\theta)$, while striving to minimize the outage probability. To find a feasible solution, we propose a pragmatic approach that sequentially tackles the following two-stage optimization problems: the minimization of outage probability and the maximization of achievable rate. 
\\
\indent
We first consider the minimization of outage probability for arbitrary MS $k$ under pure NLoS conditions, neglecting the case of any LoS. Let $\Pr\left \{\widetilde{\Gamma}_k(\bm\theta,{\textbf G})\le {\gamma_{\text{th}} }\right \}$ be the cumulative outage probability of MS $k$. Mathematically, this minimization problem can be formulated as
\begin{equation}
\label{eqn_sys_outage}
\begin{aligned}
(\mathcal P_1):\,\, &{\mathop {\mathrm {\min}} \limits _{{\bm\theta},\,{\textbf G}}}~ \Pr\left (\widetilde{\Gamma}_k(\bm\theta,{\textbf G})\le {\gamma_{\text{th}}}\right ),\quad k\in \mathcal K^{\text{NLoS}},
\\
&\;\;\mathrm {s.t.~}~\mathrm{Tr}(\textbf {G})\leq \Omega_{\text{tot}},
\\
\end{aligned}
\end{equation}
where $\gamma_{\text{th}}$ is the required SINR threshold, corresponding to the minimum acceptable rate. By the definition of the outage probability, an outage event will take place when the SINR falls below threshold $\gamma_{\mathrm{th}}$.
\\
\indent
In the second stage, we aim to find a set of optimum ${\textbf U}_k$, ${\textbf V}_k$, and ${\bm\theta}_k$ that maximize the achievable rate under the total power constraint $\Omega_{\text {tot}}$. This can be cast as the following optimization problem:
\begin{equation}
\begin{aligned}
\label{eqn_capacity_sys}
(\mathcal P_2):\, {\mathop {\mathrm {\max}} \limits _{\bm\theta,\,{\textbf G}}}\quad &\widetilde{\mathcal R}_k(\bm\theta,{\textbf G}),\quad k\in \mathcal K^{\text{NLoS}}
\\
{\mathrm {s.t.}}\quad 
&\mathrm{Tr}(\textbf {G})\leq \Omega_{\text{tot}}, \;\widetilde\Gamma_k(\bm\theta,{\textbf G})>{\gamma_{\text{th}}}, 
\end{aligned}
\end{equation}
where $\Omega_{\text{tot}}$ is the total transmit power constraint. Unless stated otherwise, we assume that objective function $\widetilde{\mathcal R}_k(\bm\theta,{\textbf G})$ per se is convex and smooth. 
Because the objective function $\widetilde{\mathcal R}_k(\bm\theta,{\textbf G})$ in the involved variables is nonconvex, the global optimum of this constrained optimization problem is hard, in general, to be found. Since the number of RIS elements is large, the AO techniques holding one variable and the other variables fixed in a cyclical fashion may require more iterations, resulting in a practically prohibitive duration for solving the optimization problem. Furthermore, it is decisive for $\mathcal P_1$ to efficiently detect the availability of dominant links, especially when a sudden blockage occurs. To facilitate the following analysis we first need to devise a statistical inference algorithm based on hypothesis testing for random blockage detection. On the basis of trustworthy blockage detection, we provide a robust beamforming design and propose an accelerated PGD algorithm to address the above non-convex optimization problem $\mathcal P_2$. More details will be given in the sequel.
\section{Blockage Detection and Indication} 
In this section, we provide a general statistical decision framework to solve the blockage detection problem. By exploiting the reciprocity of angle-domain spatial channels, we further propose a mechanism of link availability indication.
\subsection{Binary Hypothesis Testing for Random Blockage Detection}
From the perspective of statistical decision theory, we formulate the blockage detection as a binary hypothesis testing problem, and its process can be described by a three-tuple $(\mathcal H,\mathcal Y_\varepsilon, \mathscr P_{\varepsilon}) $, where $\mathcal H=\left\{{\mathcal H}_0^{\text{LoS}},{\mathcal H}_1^{\text{NLoS}}\right\}$ is a set that consists of two hypotheses: ${\mathcal H}_0^{\text{LoS}}$ is the null hypothesis of no blockage, and ${\mathcal H}_1^{\text{NLoS}}$ is the alternative hypothesis of blockage; $\mathcal Y_\varepsilon=\left\{\mathcal Y_0^{\text{LoS}},\mathcal Y_1^{\text{NLoS}}\right\}$ is the decision region with parameter $\varepsilon$, representing the disjoint sample sets $\mathcal Y_0^{\text{LoS}}$ and $\mathcal Y_1^{\text{NLoS}}$ that decide LoS or NLoS channel condition between the BS and the MS, given $\mathcal Y_0^{\text{LoS}}\cap\mathcal Y_1^{\text{NLoS}}=\varnothing$; $\mathscr P_{\varepsilon}$ is the probability distribution of the random variable $\varepsilon$ corresponding to blockages. 
\\
\indent
For the brevity of mathematical representation, we temporarily omit ${\textbf W}_k$ in the following derivation. Note that, without loss of generality, we can rewrite (\ref{eqn_Revsignal_MSa}) as a time-domain noisy observation with unknown parameter $\varepsilon$, which is given by
\begin{equation}
\label{eqn_example}
\begin{aligned}
{{\textbf{y}}}_q(\varepsilon)={{\textbf{y}}}_k(t_q)&=\overline{\textbf y}_{\text d,k}(t_q)+\overline{\textbf y}_{\text r,k}(t_q-{\tau})+{\textbf n}_k(t_q).
\end{aligned}
\end{equation}
\indent
Notice that $\overline{\textbf y}_{\text d,k}(t_q)$ and $\overline{\textbf y}_{\text r,k}(t_q-{\tau})$ represent the signals over different propagation paths. They are resolvable in terms of path delay component $\tau$, especially when considering the spatial sparsity of mmWave channels\footnote{The estimation of path delay has been well investigated in previous literature, and a detailed treatment for this issue is available in \cite{Shahmansoori18}. Generally, the time-of-arrival (ToA) detection and received signal power measurement with a high degree of accuracy can be used to efficiently distinguish different propagation paths.}. This suggests that the presence or absence of a blockage can be identified through the detection of the direct signal. 
\\
\indent
Therefore, our goal here can be equivalently transferred to preemptively detect whether a strong LoS signal $\overline{\textbf y}_{\text d,k}$ of interest is present. This can be done by distinguishing between the following two simple hypotheses \cite{Kay98}:
\begin{equation}
\label{eqn_H0}
\begin{aligned}
\mathcal H_0^{\text{LoS}}:&\;{{\textbf{y}}_q}(\varepsilon)=\overline{\textbf y}_{\text d,k}+\overline{\textbf y}_{\text r,k}+{\textbf n}_k,\; &\textit{if}\;\; \varepsilon>\varepsilon_{\text{th}},
\\
\mathcal H_1^{\text{NLoS}}:&\;{{\textbf{y}}_q}(\varepsilon)=\overline{\textbf y}_{\text r,k}+{\textbf n}_k,\; &\textit{if}\;\; \varepsilon<\varepsilon_{\text{th}},
\end{aligned}
\end{equation}
where $\varepsilon_{\text{th}}$ is a pre-defined threshold, identifying the absence or presence of the blockage. From (\ref{eqn_Revsignal_MSa}) and (\ref{eqn_Revsignal_MSb}), we can see that $\varepsilon$ is only related to the effects of the blockages but independent of the Rician factor and the fading coefficients, and thus it can be calculated as
\begin{equation}
\label{eqn_epsilon_calcu}
\begin{aligned}
\varepsilon=\frac{1}{\upsilon_{0}}\sqrt{\frac{\vert{\overline{\textbf{y}}_{\text d,k}^H}(t_q){\textbf V}_{k}{\overline{\textbf{y}}_{\text d,k}(t_q)}\vert}{\vert{\textbf H}_{\text d,k}{\textbf G}{\textbf H}_{\text d,k}^H\vert}}.
\end{aligned}
\end{equation}
\indent
Consequently, the probability mass function (PMF) of this binary hypothesis testing problem under each hypothesis can be obtained by the following form:
\begin{equation}
\label{eqn_H01}
\begin{aligned}
\mathcal H_0^{\text{LoS}}: &\;{{\mathcal Y}_{\varepsilon}} \sim f_{{\textbf Y}_q}\left(\varepsilon>\varepsilon_{\text{th}}\vert\mathcal H_0^{\text{LoS}}\right),
\\
\mathcal H_1^{\text{NLoS}}: &\;{{\mathcal Y}_{\varepsilon}} \sim f_{{\textbf Y}_q}\left(\varepsilon<\varepsilon_{\text{th}}\vert\mathcal H_1^{\text{NLoS}}\right),
\end{aligned}
\end{equation}
where $p_{{\textbf Y}_k}(\cdot)$ is the corresponding PMF of different observations ${\textbf Y}_k$ associated with $\mathcal H_0$ or $\mathcal H_1$.
\\
\indent
Given a fixed tolerable probability of false alarm $\alpha_k$, an advantageous option in practice is to maximize the probability of correct blockage detection. Upon using the NP lemma, an alternative decision rule is to leverage the “significance level” of the test for null hypothesis $\mathcal H_0$ against alternative hypothesis $\mathcal H_1$, which is equivalent to solving the following optimization problem:
\begin{equation}
\label{eqn_NP_PF}
\begin{aligned}
 \delta_{\text{NP}}({\textbf y}_q(\varepsilon))=&\mathop{\arg\max}_{\varepsilon}\quad P_{\text D}^{\text{NLoS}}(\varepsilon_{\text{th}}), \\
&\quad\; \text {s.t.} \quad\quad P_{\text{FA}}^{\text{LoS}}(\varepsilon_{\text{th}})\le \alpha ,  
\end{aligned}
\end{equation}
where $P_{\text D}^{\text{NLoS}}(\varepsilon_{\text{th}})$ is the probability of correct blockage detection; $P_{\text{FA}}^{\text{LoS}}(\varepsilon_{\text{th}})$ is the probability of false alarm, corresponding to detection threshold $\varepsilon_{\text{th}}$.
\\
\indent
During each decision epoch, we suppose that the MS directly collects $M$ samples on the basis of observation ${\textbf y}_q(\varepsilon)$ indexed on $m\in \{0,1,\dots,M-1\}$ in time. In this way, we can model the observed data by a discrete observation set ${\textbf Y}=\left\{{\textbf Y}_1,\dots,{\textbf Y}_q,\dots\right\}$, where ${\textbf Y}_q=\left[{\textbf y}_q^{(0)},\dots,{\textbf y}_q^{(m)},\dots\right]^T\in \mathbb C^{N_{t,k}\times M}$, and ${\textbf y}_q^{(m)}$ represents each possible observation \cite{Neyman92}. 
\\
\indent
From the basics of probability theory, probability of correct detection $P_{\text D}^{\text{NLoS}}(\varepsilon_{\text{th}})$ and the probability of false alarm $P_{\text{FA}}^{\text{LoS}}(\varepsilon_{\text{th}})$ can be respectively calculated as 
\begin{equation}
\label{eqn_SE_RIS}
\begin{aligned}
P_{\text D}^{\text{NLoS}}(\varepsilon_{\text{th}})&=P\left(\mathcal H_1^{\text{NLoS}}\vert\mathcal H_1^{\text{NLoS}}\right)
\\
&=\prod\limits_{l:\; \mathcal L_{l}(\varepsilon_{\text{th}})>\Lambda}\displaystyle\int_0^{\varepsilon_{\text{th}}} f_{{\textbf y}_q^{(m)}}(\varepsilon)\,d \varepsilon=\beta,
\end{aligned}
\end{equation}
and
\begin{equation}
\label{eqn_SE_RIS}
\begin{aligned}
P_{\text{FA}}^{\text{LoS}}(\varepsilon_{\text{th}})&=P\left(\mathcal H_1^{\text{NLoS}}\vert\mathcal H_0^{\text{LoS}}\right)
\\
&=\prod\limits_{l:\; \mathcal L_{l}(\varepsilon_{\text{th}})<\Lambda}\displaystyle\int_{\varepsilon_{\text{th}}}^1 f_{{\textbf y}_q^{(m)}}(\varepsilon)d \varepsilon\le\alpha,
\end{aligned}
\end{equation}
where $\alpha \in [0,1]$ is the \textit{significance level} of the test of user $k$, which can may be viewed as a transition probability of going from the LoS state to the NLoS state.
\\
\indent
Following the above formulation and derivation, the challenge now becomes to compare the best model in class $\mathcal H_1$ to the best model in $\mathcal H_0$, which can be formalized as follows. For given $P_{\text{FA}}^{\text{LoS}}=\alpha$, NP's optimal solution achieving maximum $P_{\text D}$ can be determined by the likelihood rate testing (LRT) as \cite{Kay98,Neyman92}
\begin{equation}
\label{eqn_LRT}
\begin{aligned}
\mathcal L(\varepsilon_{\text{th}})&=\frac{\displaystyle\int_0^{\varepsilon_{\text{th}}} f_{{\textbf Y}_q}\left(\varepsilon\vert\mathcal H_1^{\text{NLoS}}\right)\,d\varepsilon}{\displaystyle\int_{\varepsilon_{\text{th}}}^1 f_{{\textbf Y}_q}\left(\varepsilon\vert\mathcal H_0^{\text{LoS}}\right)\,d\varepsilon}
\\
&=\prod\limits_{l=1}^{\mathcal M}{\frac{\displaystyle\int_0^{\varepsilon_{\text{th}}} {f_{{\textbf y}_q^{(m)}}\left(\varepsilon\vert\mathcal H_1^{\text{NLoS}}\right)\,d\varepsilon}}{\displaystyle\int_{\varepsilon_{\text{th}}}^1 {f_{{\textbf y}_q^{(m)}}\left(\varepsilon\vert\mathcal H_0^{\text{LoS}}\right)\,d\varepsilon}}\mathop{\frac{>}{<}}^{\mathcal H_1^{\text{NLoS}}}_{\mathcal H_0^{\text{LoS}}}\Lambda}. 
\end{aligned}
\end{equation}
In this paper, we construct a deterministic decision rule $\delta_{\text{NP}}: {\textbf y}_q\to \{1, 0\}$ with a LRT threshold $\Lambda$. Mathematically, it can be reformulated as 
\begin{equation}
\label{eqn_SE_RIS_Pr}
\begin{aligned}
\delta_{\text {NP}}({\textbf y}_q(\varepsilon)) =
\begin{cases} 
1;\quad &\mathcal L(\varepsilon_{\text{th}})>\Lambda
\\ 
\varepsilon_{\text{th}};\quad &\mathcal L(\varepsilon_{\text{th}})=\Lambda
\\
0; \quad \quad &\mathcal L(\varepsilon_{\text{th}})<\Lambda
\end{cases},
\end{aligned}
\end{equation}
where $\delta_{\text {NP}}(\cdot)$ is the NP criterion-based statistic test operator. 
At this point, we have solved the statistical decision problem raised at the beginning, which leads to
\begin{equation}
\begin{aligned}
\label{eqn_Decision_regions}
{\mathcal Y}_0^{\text{LoS}}&\triangleq\{{\textbf y}_q\vert\delta_{\text{NP}}({\textbf y}_q (\varepsilon))=0\},
\\
{\mathcal Y}_1^{\text{NLoS}}&\triangleq\{{\textbf y}_q\vert\delta_{\text{NP}}({\textbf y}_q(\varepsilon))=1\}.
\end{aligned}
\end{equation}
Subsequent experimental results show that the NP hypothesis testing is an efficient method in practice, which facilitates sensing with low complexity to determine whether a blockage is present or not. The most significant benefit of this method is that the BS can immediately steer the dominant beam toward the RIS when a severe blockage is present.
\subsection{Link Availability Indication with Angle Reciprocity}
\begin{figure}[!t]
\centering
\includegraphics[width=5.5cm]{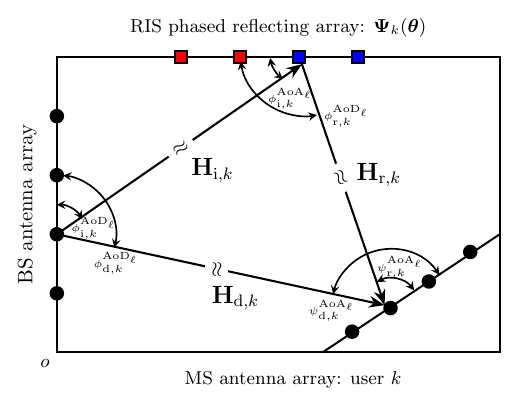}
\caption{Pictorial illustration of geometrical representation of the channel reciprocity on the azimuthal plane: Reflecting geometry satisfies certain constraints imposed by the array geometry between BS-MS and BS-RIS-MS channels, where $\vartheta$ and $\theta$ denote the spatial AoA and AoD, respectively.}
\label{figure}
\end{figure}
For the studied system in this paper, it is worth noting that the angle-dependent characteristic of RIS has been verified through full-wave simulations\cite{Chen20}. Although the artificial reflection properties of RIS do not satisfy the reflection theorem, the angle reciprocity still holds in RIS-aided mmWave systems since the BS and RIS are generally fixed and mmWave is capable of providing the requisite for the spatial discrimination in the angular domain \cite{Shahmansoori18}. Based on these facts, one important finding in this paper is that the angle reciprocity is retained in reverse links, e.g., BS-MS links, RIS-MS links, and BS-RIS links. Geometrically, this suggests that any pair of angles of incidence and reflection is interdependent, such that the angle reciprocity uniformly holds on both azimuth (horizontal) and elevation (vertical) projection planes without explicitly calculating CSI in the opposite direction. Fig. 2 graphically depicts the angle-dependent relations among the BS, MS, and RIS.
\\
\indent
With this far-field aerial view, denoting ${\mathcal A}^{-1}$ as an inverse operator of the angle transformation and exploiting the aforementioned channel reciprocity, the mapping relation of the angle reciprocity from AoA to AoD can be explicitly expressed as
\begin{equation}
\begin{aligned}
\label{eqn_recip_1}
{\psi}^{\text{AoD}}_{k}  \sim{{\mathcal A}^{-1}}( \measuredangle{{\psi}^{\text{AoA}}_{k}};\measuredangle{{(\phi,\varphi)}^{\text{AoA}}_{k}}),~\forall \, k.
\end{aligned}
\end{equation}
\indent
To fully exploit the advantages of the proposed blockage detection approach, we further propose a mechanism of link availability indication. Specifically, the BS transmits pilot symbols to the MS for channel estimation, and the MS simultaneously performs blockage detection and sends feedback on the link availability through dedicated feedback channels. Leveraging the angle reciprocity revealed in Fig. 2, a complex channel estimation procedure can be avoided since the location of the MS is prone to be determined by appropriate positioning techniques\cite{Shahmansoori18}. In this case, even if the acquisition of CSI at the BS is partial, the BS is capable of reconstructing the requisite AoD through the AoA measurement, and vice versa. Based on the blockage detection results provided by the MS, the BS can easily decide whether or not to establish an alternative reflection link. For uncertain BS-RIS-MS links, the BS is capable of proactively restoring unknown or inaccurate angle information by means of angle reciprocity, thereby effectively mitigating the impact of insufficient signal processing capabilities, due to the passive nature of the RIS.
\\
\indent
Through the aforementioned methods, we can solve the blockage detection problem on an accurate basis. From the reliability perspective, the proposed blockage-aware approach focuses on preemptively excluding potential outages caused by random blockages. As a consequence, the system outage performance and transmission reliability can thus be improved. 
\section{Outage Minimization and Rate Maximization} 
In this section, we focus on solving previously formulated optimization problems $\mathcal P_1$ and $\mathcal P_2$. It should be emphasized that both optimization problems are resolved on the premise that the potential link blockages can be accurately distinguished by means of the methodology presented in Section III. In this case, since the RIS-MS reflection link can take effect immediately, the BS has the opportunity to preventively steer the beam toward the RIS, thereby avoiding the worst-case system outage.
\subsection{Outage Minimization via Robust Beamforming}
It is worth pointing out that if the blockages can be efficiently detected, the worst-case system outage can be substantially circumvented. As a result, the system outage probability will be considerably reduced. It suggests that originally formulated problem $\mathcal P_1$ has been roughly solved. In this subsection, we focus on solving the outage-constrained robust beamforming optimization problem, including jointly tailoring transmit precoders at the BS and the reflective PSs at the RIS. To combat the uncertainty of the cascaded BS-RIS-MS channel, we further devise a beamwidth-variant control scheme, which is capable of further reducing the outage probability and providing more resilient connectivity when the dominant LoS link is blocked.
\\
\indent
In essence, for many RIS aided communication applications, such passive reflecting surfaces are required to intercept as much radiated energy as possible from a far-field source, and to generate the reflected EMs with a sufficiently large directional gain. Also, it is proven that the received signal power is proportional to the size of the RIS surface \cite{Ozdogan20}. For mmWave communications, it suggests that if more energy is to be harvested through the RIS, it is more likely for the BS to choose a narrower redirect beam toward the RIS. Otherwise, according to the law of conservation of energy, a wider beam necessarily results in a more dispersed radiation distribution, such that most of the energy from the far-field BS misses the receiving aperture of the RIS. An in-depth discussion of the far-field properties of RISs can be found in \cite{Ozdogan20}.
\\
\indent
Motivated by the above elaboration, we propose a beamwidth-variant beamforming solution enabling flexible beamwidth adaptation. Normally, the BS-RIS beam is designed to be with the narrowest beamwidth, such that the optimization of transmit precoder ${\textbf F}_k$ and PSM $\bm\Psi(\bm\theta)$ can be decoupled. As a consequence, the solution to the outage-constrained problem $\mathcal P_1$ can be divided into the following two steps. The narrowest BS-MIS beam is designed first. Then, assuming a local optimum ${\textbf F}_k^*$ has been obtained, an adaptive beamwidth control method can be applied to further decrease the outage probability.
\\
\indent
The first step of the algorithm considers the design of hybrid precoder ${\textbf F}_{\text{BB}_k}$ and ${\textbf F}_{\text{RF}_k}$, with the purpose to minimize the beamwidth of the main lobe. Most commonly, the beamwidth can be defined by the half-power beamwidth (HPBW), equivalent to the angular width of the radiation pattern that is 3-dB lower than the maximum gain of the beam, i.e., the beam peak \cite{Sun18}. On the azimuthal plane, assume that the boresight beam peak is at steering angle $\upphi$. The beamwidth can be defined as difference $\vert\upphi^{+}_{3\text{-}\text{dB}}-\upphi^{-}_{3\text{-}\text{dB}}\vert$, where $\upphi^{+}_{3\text{-}\text{dB}}$ and $\upphi^{-}_{3\text{-}\text{dB}}$ correspond to the 3-dB angles at the ambilateral direction of the beam peak, respectively. Thus, the design problem of the narrowest transmit beam can be written as 
\begin{equation}
\label{eqn_bs_beamwidth}
\begin{aligned}
\mathop{\arg\min}\limits_{{\textbf F}_{\text{BB}_k},\,{\textbf F}_{\text{RF}_k}}{\Phi}_{\text{BS}}({\textbf F}_{\text{BB}_k},{\textbf F}_{\text{RF}_k}):=\vert\upphi^{+}_{3\text{-}\text{dB}}-\upphi^{-}_{3\text{-}\text{dB}}\vert.
\end{aligned}
\end{equation}
In general, narrowest possible beamwidth ${\Phi}^{\text{min}}_{\text{BS}}({\textbf F}_{\text{BB}_k},{\textbf F}_{\text{RF}_k})$ can be determined by optimal beamforming vector ${\textbf F}_{\text{BB}_k}$ and ${\textbf F}_{\text{RF}_k}$ in a hierarchical multi-resolution codebook, where the narrowest beamwidth is associated with the highest level of codebook. In terms of controlling variable beamwidth, a pioneer contribution to the design of HAD codebooks can be found in \cite{Alkhateeb14-2}.
\\
\indent
Meanwhile, it is confirmed that a proper beamwidth plays a vital role in tracking mobile receivers \cite{Alkhateeb14}. Specifically, the beamwidth required to support the mobile access scenarios is slightly different. It is noteworthy that the narrower beamwidth demands more precise beam alignment due to the movement of receivers, while frequent beam alignment may entail considerable overhead. Furthermore, the narrow beams are more sensitive to angular changes, because of which the effects of imperfect CSI will be prominent, since the misaligned beams significantly reduce the signal gain at the receiver, incurring an increased outage probability. Obviously, the aforementioned facts still hold for the RIS-aided mmWave access. With regard to RIS reflective surfaces, it is known that the beamwidth of RIS reduces as the number of reflective elements increases.
\\
\indent
Thus, in the followoing passive beamforming design, we formulate an adjustable beamwidth control strategy by scaling the number of the RIS reflective elements. In a more compact form, (\ref{eqn_sys_outage}) can be rewritten as
\begin{equation}
\label{eqn_RIS_beamwidth_outage}
\begin{aligned}
{\mathop {\mathrm {\min}} \limits _{{\Phi_{\text{RIS}}({\bm\theta},\mathcal N_k)},\,{\textbf G}}}~ &\Pr\left (\widetilde\Gamma_{k}(\bm\theta,{\textbf G})\le {\gamma_{\text{th}}}{\vert_{{\Phi_{\text{RIS}}({\bm\theta}},\mathcal N_k)}}\right ),\, k\in \mathcal K^{\text{NLoS}},
\\
\mathrm {s.t.}\quad&\mathrm{Tr}(\textbf {G})\leq \Omega_{\text{tot}},\quad{\mathrm{MSE}}_{\bm\theta}<\zeta_{\bm\theta}^{\text{max}},
\end{aligned}
\end{equation}
where $\Phi_{\text{RIS}}({\bm\theta},\mathcal N_k)$ is a function of ${\bm\theta}$, and $\mathcal N_k$, controlling the beamwidth of the RIS; $\zeta_{\bm\theta}^{\text{max}}$ is the maximum tolerable level of inaccurate PS estimates. $\Phi_{\text{RIS}}({\bm\theta},\mathcal N_k)$ is inversely proportional to the distance between the RIS and the MS.
\\
\indent
Compared to a constant-beamwidth strategy, the adjustable beamwidth allows for more flexibility by scaling $\mathcal N_k$, and there exists an optimal beamwidth that significantly suppresses the outage probability. It is shown in the simulation part of this paper that even though a broadened beam may temporarily reduce the achievable rate of the blocked user, the downlink user sum rate roughly remains constant. The intrinsic reason for this is that a wider beam requires less training overhead, especially with a significant reduction in outage probability. Because channel estimation errors are inevitable in practice, the stringent accuracy requirements can be relaxed since the BS can readily reconstruct the indirect RIS-MS channel matrix by leveraging the angle reciprocity revealed in Fig. 2. In this case, the CSI estimation error at the BS can be regarded as acceptable as long as the constraint of the maximum tolerance level $\zeta_{\bm\theta}^{\text{max}}$ is satisfied. 
\subsection{Rate Maximization via Projected Gradient Descent} 
We now consider the globally optimal solution to problem $\mathcal P_2$. In general, a multivariate optimization problem similar to problem $\mathcal P_2$ can be straightforwardly tackled by the conventional AO technique, the main idea of which is that each variable is optimized in an alternating manner while the others are fixed. However, the AO-based algorithm might not be applicable to RIS-aided mmWave communication systems. An underlying reason is that an RIS consists of massive passive elements, whereby the optimization of problem $\mathcal P_2$ requires a large number of adjustments of the PSs. This would, of necessity, involve a large number of optimization variables, and thus the computational complexity becomes prohibitive. Besides, the objective function ${\widetilde{\mathcal R}}_k$ in (14) is still nonconvex due to the inequality constraints, which also makes problem $\mathcal P_2$ difficult to solve. Among optimization algorithms, the gradient descent method is a vital tool, and provably better for finding the local minima of an objective function in the steepest descent direction.
\\
\indent
To this end, we devise an accelerated PGD algorithm with Nesterov's momentum. From the extreme value theorem, even though objective function $\widetilde{\mathcal R}$ is nonconvex, its constituents can be supposed to be a series of locally convex and continuously differentiable functions, which can be regarded as a manifold with the Lipschitz continuous gradient that can be efficiently solved by numerical schemes. By applying the PGD algorithm, a local extremum, i.e., $\widetilde{\mathcal R}^*(\bm\theta_k,{\textbf G}_k): {\mathbb R}^n\to \mathbb R$, as well as an absolutely global maximum $\widetilde{\mathcal R}^{\star}(\bm\theta_k,{\textbf G}_k)$ can be found recursively. In what follows, we propose a three-stage approach to implement this accelerated version of PGD. 
\\
\indent
{\textit{1) Solve constrained PGD with projection operation}}: By applying the constrained optimality theorem, the optimization of $\mathcal P_2$ turns out to find the closest ${\bm\theta}^{\star}$ and ${\textbf G}^{\star}$ in a feasible set $\mathcal F=\{\Theta,\mathcal G\}, \mathcal F \subset \mathbb R^n$ to maximize objective function $\widetilde{\mathcal R}$ \cite{Bertsekas99}. For proposed problem $\mathcal P_2$, $\Theta$ and $\textbf G$ are respectively given by $\Theta:=\{\bm\theta \in {\mathbb R}^n\vert \widetilde\Gamma_k(\bm\theta,{\textbf G})>{\gamma_{\text{th}}}\}$ and ${\mathcal G}:=\{\textbf G \in {\mathbb R}^n\vert \mathrm{Tr}(\textbf {G})\leq \Omega_t\}$ that both of the set of points satisfy the inequality constraints.
\\
\indent
To solve problem $\mathcal P_2$  in a standard gradient descent fashion, we can reformulate, in terms of the feasible set $\mathcal F$, the optimization problem in a more compact form as follows:
\begin{equation}
\begin{aligned}
\label{eqn_M_objective}
{\mathop {\arg\min}  \limits_{{\bm\theta}^{\star} \in \Theta,\; {\textbf G}^{\star} \in \mathcal G}}\quad &{\widetilde{\mathcal R}}_k^{-1}(\bm\theta,{\textbf G}),\quad k \in {\mathcal K}^{\text{NLoS}}.
\end{aligned}
\end{equation}
Similarly, objective function ${\widetilde{\mathcal R}}_k^{-1}$ holds continuously differentiable and has a known gradient Lipschitz constant e.g., $L$-Lipschitz continuous. Our main focus of this subsection is to find the global minima of function $\mathcal R_k^{-1}$, which is equivalent to excluding all local minima. 
\\
\indent
In the case of RIS-aided mobile mmWave communications, a key fact is that the spatial channel can be equivalently represented in the beamspace domain, i.e., the angular domain. For the beamspace in $\mathbb R^2$, we just need to consider the azimuth and elevation angles during the beam search. In addition, this fact also suggests that to retain stable connectivity, the RIS reflected beam corresponding to feasible set $\mathcal F$ is necessarily associated with the motion trajectory of the MS under the NLoS condition \cite{yang202}. By leveraging these facts, we can represent whole set ${\mathcal F}$ as a set of mutually independent subsets whose collections are used to represent the continuous-time approximation of motion trajectory. Under this reasonable assumption, for each  motion trajectory $w$, where $w=1,2,\dots,W$, there exists ${\mathcal F}={\{{\mathcal F}_{1},{\mathcal F}_{2},\dots,{\mathcal F}_{W}\}}$. That is to say, for any subset $\mathcal F_w \subset \mathcal F$, $\{{\mathcal F}_{w}\}_{w=1}^W$ is a spanning set for $\mathcal F$. By analogy, for given feasible sets $\Theta=\{\Theta_1,\Theta_2,\dots,\Theta_W\}$ and $\mathcal G=\{\mathcal G_1,\mathcal G_2,\cdots,\mathcal G_W\}$, the projection operation is correspondingly converted to the subspace projection along the motion trajectory, with the goal of finding the ``closest`` subset to ${\mathcal F}_w$ in the sense of Euclidean norm. From this the above reasoning, feasible subset $\mathcal F_w$ is spanned by
\begin{equation}
\begin{aligned}
\label{eqn_GM_distribution}
{{\Theta}_w}&={\textrm{Span}({\bm\theta}_w)}\,={\textrm{Span}({\theta}_{w,1},{\theta}_{w,2},\dots,{\theta}_{w,\imath\jmath})},
\\
{{\mathcal G}_w}&={\textrm{Span}({\textbf G}_w)}={\textrm{Span}({\textbf g}_{w,1},{\textbf g}_{w,2},\dots,{\textbf g}_{w,N_s})},
\end{aligned}
\end{equation}
where ${\textbf g}_{w,N_s}\in \mathbb C^{N_s}$ is a row vector of ${\textbf G}_w$.
\\
\indent
We can define ${\textbf P}_{\mathcal F_w}(\textbf u)=\frac{\textbf u\cdot\textbf v}{\vert\vert \textbf v \vert\vert^2 }\textbf v$ as the Euclidean projection from vector $\textbf u$ onto vector $\textbf v$ in $\mathcal F_w$. Exactly speaking, it refers to a mapping relation from $\mathbb R^n$ to $\mathbb R^n$, which in our case is to simply find ${\bm\theta}_w$ and ${\textbf G}_w$ that are closest to $\Theta_w$ and $\mathcal G_w$, respectively. Mathematically, the projection operation onto constraint set $\Theta_w$ and $\mathcal G_w$ can be cast as the following optimization problems:
\begin{subequations}
\begin{align}
\label{eqn_GM_distribution}
{\textbf P}_{\Theta_w}({\bm\theta}_w)&=\arg\min_{\bm\vartheta\in \Theta_w}~\vert\vert\bm\vartheta-{\bm\theta}_w\vert\vert_2,
\\
{\textbf P}_{\mathcal G_w}({\textbf G}_w)&=\arg\min_{{\textbf Q}\in \mathcal G_w}~\vert\vert{\textbf Q}-{\textbf G}_w\vert\vert_2.
\end{align}
\end{subequations}
The key idea of the above method is to find suitable vectors ${\bm\theta}_w$ and ${\textbf G}_w$ within a smaller subspace $\mathcal F_w$. By projecting onto the constraint set, constrained optimization problem $\mathcal P_2$ can be fixed through a simple modification of standard gradient descent. Furthermore, since the PGD algorithm does not need to search the whole space, the computational complexity is thereby reduced. 
\\
\indent
Following the above reasoning and manipulation, we devise an accelerated PGD algorithm in the next stage. The prestigious Nesterov’s momentum method is applied to circumvent local minima, especially to prevent the PGD algorithm from getting stuck around saddle points of $\widetilde{\mathcal R}_k$. 
\\
\textit{2) Accelerate PGD with Nesterov's momentum}:
In standard Nesterov's accelerated gradient method, an extra momentum term is first introduced to the gradient descent on $\widetilde{\mathcal R}_k^{-1}$ at each iteration. After that, parameters ${\bm\theta}_w$ and ${\textbf G}_w$ are iteratively updated by re-projecting each of them onto constraint set $\Theta_w$ and $\mathcal G_w$. The updating process of Nesterov's momentum and parameters over two successive steps is summarized through the following recursive formulas \cite{Nesterov04}:
\begin{equation}
\label{eqn_Proj_theta}
\begin{aligned}
{\bm\eta}_w^{(n+1)}&={\bm\nu}{\bm\eta}_w^{(n)}-{\mu}_{{\bm\theta}_w}^{(n)}\nabla_{{\bm\theta}_w}\widetilde{\mathcal R}_k^{-1}\left(\bm\theta_w^{(n)},{\textbf G}_w^{(n)}\right),
\\
{{\bm\theta}}^{(n+1)}_w&={\textbf P}_{\Theta_w}\left({\bm\theta}_w^{(n)}+{\bm\eta}_w^{(n+1)}\right),
\end{aligned}
\end{equation}
and
\begin{equation}
\label{eqn_Proj_G}
\begin{aligned}
{\bm{\chi}}_w^{(n+1)}&={\bm\Delta}{\bm{\chi}}_w^{(n)}-{{\mu}}_{{\textbf G}_w}^{(n)}\nabla_{{\textbf G}_w}\widetilde{\mathcal R}^{-1}_k\left(\bm\theta_w^{(n)},{\textbf G}_w^{(n)}\right),
\\
{{\textbf G}}^{(n+1)}_w&={\textbf P}_{{\mathcal G_w}}\left({\textbf G}_w^{(n)}+{\bm{\chi}}_w^{(n+1)}\right),
\end{aligned}
\end{equation}
where ${\bm\eta}_w^{(n+1)}$ and ${\bm{\chi}}_w^{(n+1)}$ are the accumulated momentum terms at the $(n+1)$th iteration; hyperparameter $\bm\nu\in[0,1]$ and $\bm\Delta\in[0,1]$ represent the level of inertia in the descent direction (a.k.a the momentum coefficients), which can be usually chosen by trial and/or with a model selection criterion \cite{Bengio00}; $\mu_{{\bm\theta}_w}^{(n)}\in (0,1)$ and $\mu_{{\textbf G}_w}^{(n)}\in (0,1)$ are the step sizes used in $n$th iteration.
\\
\indent
Directly following from the derivation in \cite{Perovic21}, the complex-value gradient of $\widetilde{\mathcal R}^{-1}_k$ with respect to ${\bm\theta}_w$ and ${\textbf G}_w$ can be explicitly computed as follows:
\begin{subequations}
\label{eqn_EM_Bayesian}
\begin{align}
\nabla_{{\bm\theta}_w}\widetilde{\mathcal R}^{-1}_k({\bm\theta}_w,{\textbf G}_w)&=\mathcal {V}_d\left({\textbf H}_{\text{RU},k}^H{\bm \Sigma}^{-1}{\textbf G}({\bm\theta}_w){{\textbf H}}_{\text{BR},k}\right),
\\
\nabla_{{\textbf G}_w}\widetilde{\mathcal R}^{-1}_k({\bm\theta}_w,{\textbf G}_w)&={\textbf G}^H({\bm\theta}_w){\bm \Sigma}^{-1}{\textbf G}({\bm\theta}_w),
\end{align}
\end{subequations}
where $\mathcal V(\cdot)$ denotes the vectorization operator that stacks the columns to create a single long column vector. The detailed calculation procedure of the complex-value gradient is rigorously derived in \cite{Perovic21}. Without repeating the procedure, we choose to omit these derivation details for brevity, and interested readers are referred to \cite{Perovic21} for a complete exposition. 
\\
\indent
From a physics perspective, momentum terms ${\bm\eta}_w^{(n+1)}$ and ${\bm{\chi}}_w^{(n+1)}$ are equivalent to the momentum of inertia the  “average” of the sequence ${\bm\eta}_w^{(n)}$ and ${\bm{\chi}}_w^{(n)}$ after $n$ iterations. Particularly, since this momentum-aided PGD method makes full use of stored knowledge about the gradient of objective function $\widetilde{\mathcal R}_k^{-1}$ at previous descent directions. It has been proven that this learning-based method is capable of better estimating the next descent step \cite{Bengio00}, thereby helping to eventually escape the potential saddle points of the objective function.
\\
\textit{3) Determine an appropriate step size}:
To guarantee a higher convergence rate, the choice of step size is critical to the PGD algorithm. However, a drawback of Nesterov's method is that the convexity parameter of the objective function is required to compute an appropriate step size. Basically, the strict convexity of $\widetilde{\mathcal R}_k$ is hard to guarantee in each step of the descent process of the gradient since there are different degrees of convexity. To solve this dilemma, our focus in the sequel is to apply a backtracking line search in accordance with the Armijo rule to determine an appropriate step size. 
\\
\indent
In this approach, the step size is determined in terms of the criteria of sufficient decrease and does not require prior knowledge of convexity parameters. We first compute a search direction and then find an acceptable step size with no additional gradient evaluation. Then, a suitable step size can be chosen in each iteration $n$ such that the dynamics of the gradient can be calculated. For ease of notation, we define ${\bm\xi}^{(n)}_{{\bm\theta}_w}=-\nabla_{{\bm\theta}_w}\widetilde{\mathcal R}_k^{-1}\left(\bm\theta_w^{(n)},{\textbf G}_w^{(n)}\right)$ and ${\bm\Xi}^{(n)}_{{\textbf G}_w}=-\nabla_{{\textbf G}_w}\widetilde{\mathcal R}^{-1}_k\left(\bm\theta_w^{(n)},{\textbf G}_w^{(n)}\right)$ to be the search direction with respect to $\bm\theta_w^{(n)}$ and ${\textbf G}_w^{(n)}$ at the $n$th iteration, respectively. 
\\
\indent
Letting $\beta_k\in (0,1)$ denote the constant discount factor, the backtracking line search for choosing a proper step size at the $n$th iteration is to perform pullbacks by iteratively updating $\mu_{{\bm\theta}_w}^{(n)}=\beta_{{\bm\theta}_w}^{m_k}$ and $\mu_{{\textbf G}_w}^{(n)}=\beta_{{\textbf G}_w}^{m_k}$ with the smallest integer $m_k$. More explicitly, it is equivalent to solving the following minimization problems: 
\begin{equation}
\begin{aligned}
\label{eqn_linesearch_stepsize}
&\mathop{\arg\min} \limits_{m_k} \widetilde{\mathcal R}^{-1}_k\left({\textbf P}_{\Theta_w}\left(\bm\theta_{w}^{(n)}+\beta_{{\bm\theta}_w}^{m_k}{\bm\xi}^{(n)}_{{\bm\theta}_w}+{\bm\nu}{\bm\eta}_w^{(n)}\right)\right), 
\\
&\mathop{\arg\min} \limits_{m_k} \widetilde{\mathcal R}^{-1}_k\left({\textbf P}_{{\mathcal G}_w}\left({\textbf G}_{w}^{(n)}+\beta_{{\textbf G}_w}^{m_k}{\bm\Xi}^{(n)}_{{\textbf G}_w}+{\bm\Delta}{\bm{\chi}}_w^{(n)}\right)\right). 
\end{aligned}
\end{equation}
Starting with $m_k = 0$ and then incrementing $m_k$ until the following optimality conditions (sufficient decrease) are satisfied: 
\begin{equation}
\begin{aligned}
\label{eqn_GM_distribution}
\widetilde{\mathcal R}^{-1}_k\left({\bm\theta}_w^{(n)}\right)-\widetilde{\mathcal R}_k^{-1}\left({\bm\theta}_w^{(n+1)}\right)&\ge\uppi_{{\bm\theta}_w}\beta_{{\bm\theta}_w}^{m_k} \vert\vert{\bm\xi}^{(n)}_{{\bm\theta}_w}\vert\vert^2,
\\
\widetilde{\mathcal R}^{-1}_k\left({\textbf G}_w^{(n)}\right)-\widetilde{\mathcal R}_k^{-1}\left({\textbf G}_w^{(n+1)}\right)&\ge\uppi_{{\textbf G}_w}\beta_{{\textbf G}_w}^{m_k} \vert\vert{\bm\xi}^{(n)}_{{\textbf G}_w}\vert\vert^2.
\end{aligned}
\end{equation}
Here, parameters $\uppi_{{\bm\theta}_w} \in (0,1)$ and $\uppi_{{\textbf G}_w} \in (0,1)$ control the acceptable error tolerance of the line search procedure. To achieve local minima rapidly, larger step size $\mu_k$ is generally preferable for fast descent at current point $\widetilde{\mathcal R}^{-1}_k(\bm\theta,{\textbf G})$. In practice, there is a balance that must be maintained between taking $\mu_k$ as large as possible and not having to evaluate the function at many points. Such a balance can be obtained with an appropriate selection of parameters $\beta$ and $\upsilon_k$, such that saddles point can be avoided.
\\
\indent
Now we are ready to directly solve problem ${\mathcal P}_2$. Specifically, we devise an iterative algorithm, of which the pseudo codes are summarized in Algorithm 1. In general, the initial point $\{{\bm\theta}_w^{(0)},{\textbf G}_w^{(0)}\}$ can be randomly initialized or artificially set by guessing the location of a local minimum. In the standard Nesterov momentum updating step (line 10), hyperparameters $\bm\nu$ and $\bm\Delta$ are applied to preserve the historical gradient information that can produce new gradient values by an arbitrary number of training steps. With the Nestorov momentum, the size of the next-step gradient can be foreseeingly adjusted, thereby enabling the PGD algorithm to achieve faster convergence in a more responsive way. To simplify notation, we use ${\textbf{grad}}_{{\bm\theta}_w^{(n)}}$ and ${\textbf{grad}}_{{\textbf G}_w^{(n)}}$ to represent the corresponding gradient vectors. Through both, we can stipulate a termination criterion: the average gradient is sufficiently small after a certain number of iterations in comparison with a prescribed threshold, i.e., $\vert\vert{\textbf{grad}}_{{\bm\theta}_w^{(n)}}-{\textbf{grad}}_{{\bm\theta}_w^{(n-1)}}\vert\vert<\epsilon$. Therefore, once a stationary point is reached, i.e., the algorithm meets the preset optimality conditions, this optimization procedure will be terminated, and the stationary point will be regarded as the global minimum. 
\begin{algorithm}[!t]
\caption{Solution of projected gradient descent (PGD).}
\textbf{Input:} $\{{\bm\Theta}_w\}_{w=1}^W,\{{\mathcal G}_w\}_{w=1}^W$\Comment{non-empty feasible sets}\\
\begin{algorithmic}[1]
\\ 
\textbf{Initialization:} ${\bm\theta}_w^{(0)},{\textbf G}_w^{(0)},{\bm\nu}\leftarrow 0,{\bm\Delta}\leftarrow 0, \mu_{{\bm\theta}_w}^{(0)}, \mu_{{\bm\theta}_w}^{(0)}$ 
\For {$n=0,1,\dots$}
\\
\hspace*{1em}Compute gradient ${\textbf{grad}}_{{\bm\theta}_w^{(n)}}$ and ${\textbf{grad}}_{{\textbf G}_w^{(n)}}$ by (34)\\
\hspace*{1em}Accelerate ${\bm\eta}_w^{(n+1)}={\bm\nu}{\bm\eta}_w^{(n)}-{\mu}_{{\bm\theta}_w}^{(n)}{\textbf{grad}}_{{\bm\theta}_w^{(n)}}$\\
\hspace*{5.65em}${\bm{\chi}}_w^{(n+1)}={\bm\Delta}{\bm{\chi}}_w^{(n)}-{{\mu}}_{{\textbf G}_w}^{(n)}{\textbf{grad}}_{{\textbf G}_w^{(n)}}$
\\
\hspace*{1em}Update ${\bm\theta}_w^{(n+1)}\leftarrow{\textbf P}_{\Theta_w}\left({\bm\theta}_w^{(n)}+{\bm\eta}_w^{(n+1)}\right)$
\\
\hspace*{4.15em}${{\textbf G}}^{(n+1)}_w\leftarrow{\textbf P}_{{\mathcal G_w}}\left({\textbf G}_w^{(n)}+{\bm{\chi}}_w^{(n+1)}\right)$\Comment{projection}
\\
\hspace*{1em}Try step sizes $\mu_{{\bm\theta}_w}^{(n)}:=1,\beta_{{\bm\theta}_w},\beta_{{\bm\theta}_w}^{2},\dots,$\\
\hspace*{7em}$\mu_{{\textbf G}_w}^{(n)}:=1,\beta_{{\textbf G}_w},\beta_{{\textbf G}_w}^{2},\dots,$\Comment{line searching}
\\
\hspace*{1em}Update current Nesterov's momentum ${\bm\nu},{\bm\Delta}$ [33]\\
\hspace*{1em}Repeat until $\vert\vert {\textbf{grad}}_{{\bm\theta}_w^{(n)}}-{\textbf{grad}}_{{\bm\theta}_w^{(n-1)}}\vert\vert^2<\epsilon$
\\
\textbf{Output:} $\bm\theta^{*},{\textbf G}^{*}$
\EndFor
\end{algorithmic}
\end{algorithm}
\subsection{Analysis of Computational Complexity and Convergence}
For the proposed PGD optimization algorithm, the overall computational complexity mainly depends upon the product of the number of required iterations $I_{\text{PGD}}$ and the amount of computation required for the projection operations $N_{\text{proj}}$ as well as the gradient calculations $N_{\text{grad}}$ in each iteration. For any closed convex set $\mathcal F \subset \mathbb R^n$, the computation of projection operations is dominated by the computational overhead of a Euclidean projection on simplex convex set $\{\Theta,\mathcal G\}$. Correspondingly, the standard Euclidean norm (cf. (35a) and (35b)) that can be solved efficiently is required to take $N_{\text{proj}}={\mathcal O}(({\mathcal N}_k+N_{t,k}^2N_r^2)\log ({\mathcal N}_k+N_{t,k}^2N_r^2))$ floating-point operations in general. The computational complexity of the complex-value gradient is approximated by $N_{\text{grad}}=\mathcal O(\mathcal N_kN_{t,k}N_r)$ on condition that $\mathcal N_k\gg N_{t,k}$. It is evident that the overall computational cost of Algorithm 1 is approximate as $I_{\text{PGD}}\times(N_{\text{proj}}+N_{\text{grad}})$. 
\\
\indent
The convergence rate of Algorithm 1 can be improved by decoupling the objective function. According to the inference in Section III-A, when the BS-MS links are unavailable, the BS will generate the narrowest BS-RIS beam, in the sense that hybrid precoders ${\textbf F}_{{\text {BB}}_k}$ and ${\textbf F}_{{\text {RF}}_k}$ can be viewed as relatively constant or vary only within a small range. Combined with this mechanism, it is feasible to substantially decouple the optimization for objective function $\mathcal R_k$. To be specific, we can first complete the optimization of ${\textbf G}_w$, which approximately stays invariant throughout the algorithm. In this way, the optimization of ${\bm\theta}_w$ can be independently proceeded. The advantage of this approach is that even though this strategy is suboptimal, the optimization of problem $\mathcal P_2$ can be configured as a simplex-constrained one-dimensional function, such that each iteration will come with a much lower computational burden. In the subsequent simulation, we examine the convergence of the overall algorithm and provide evidence that the accelerated PGD algorithm converges faster than the baseline algorithms. 
\section{ Numerical Results}

\begin{table}[!t]
\caption{System parameters and simulation settings.}
\label{tab:param_simul}
\begin{center}
\begin{tabular}{llll}
\toprule
System Parameter& Simualtion setting\\ \hline \hline
Carrier frequency &28 GHz\\
Bandwidth &100 MHz\\
Cell radius & 400 m\\ 
Channel model & Rician fading channel model\\
Path loss & FSPL ratio \cite{Perovic21}\\
Modulation format &QPSK\\
Receiver noise figure&5 dB \\
$N_t$&32\\
$N_r$&4\\
$\mathcal N (\mathcal I\times\mathcal J)$& 2048 (32$\times$64) \\
\bottomrule
\end{tabular}
\end{center}
\end{table}

In this section, we evaluate the outage probability and achievable rate with extensive Monte Carlo simulations. We consider a rectangle RIS comprising 2,048 elements arranged in a $32\times64$ matrix. The center of ULA of the BS and the center of the RIS panel are assumed to be in the same horizontal plane and have the same height, and the position of the
RIS is fixed. Without loss of generality, the distance between BS and RIS $d_{\text{BS-RIS}}$ is set to 100 meters, and the corresponding angle between BS and RIS is fixed at 45 degrees. To evaluate the path loss of direct BS-MS link and cascaded BS-RIS-MS link, we adopt the total free space path loss (FSPL) ratio model proposed by \cite{Perovic21}. The system is assumed to operate at the 28 GHz carrier frequency, and the other key simulation setups are given in Table 1.
\subsection{Performance of Blockage Detection}
In what follows, our first goal is to investigate the impact of link blockage $\varepsilon_k$. To simulate random blockage events for the downlink, we adopt a probabilistic model [15], [16], by which each path undergoes a random and independent blockage. Specifically, we utilize the homogeneous Poisson process to emulate the statistical properties of blockage behavior of mmWave communication links between the BS and the MS. To be more specific, random $\varepsilon_k$ signifies the effect of different blockages, and the sudden blockage event is modeled using a two-dimensional homogeneous Poisson process. 
\begin{figure}[!t]
\centering
\begin{minipage}[t]{0.48\textwidth}
\centering
\includegraphics[width=8.5cm]{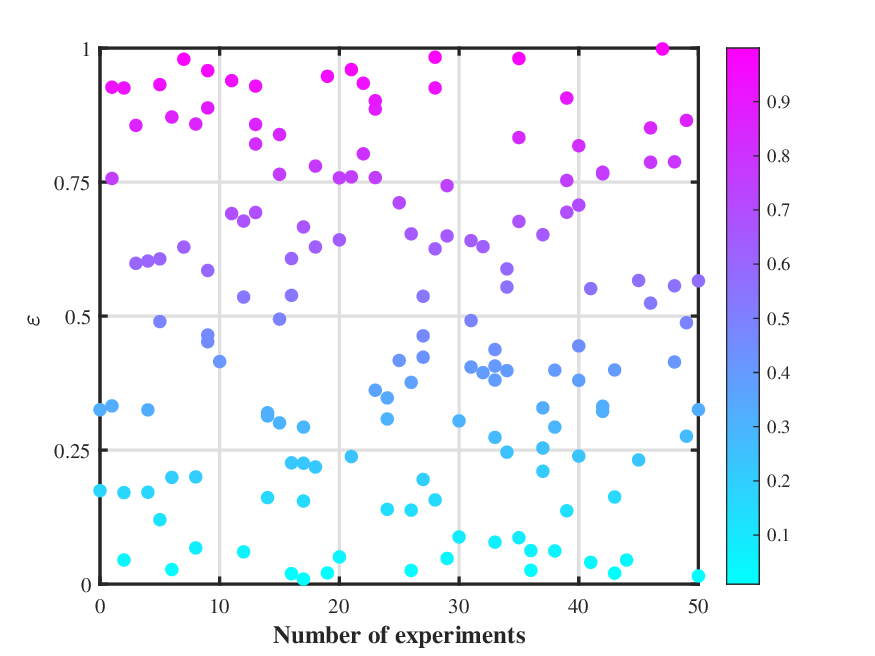}
\caption{Distribution of the blockages: An example for illustration.}
\end{minipage}
\begin{minipage}[t]{0.48\textwidth}
\centering
\includegraphics[width=8.5cm]{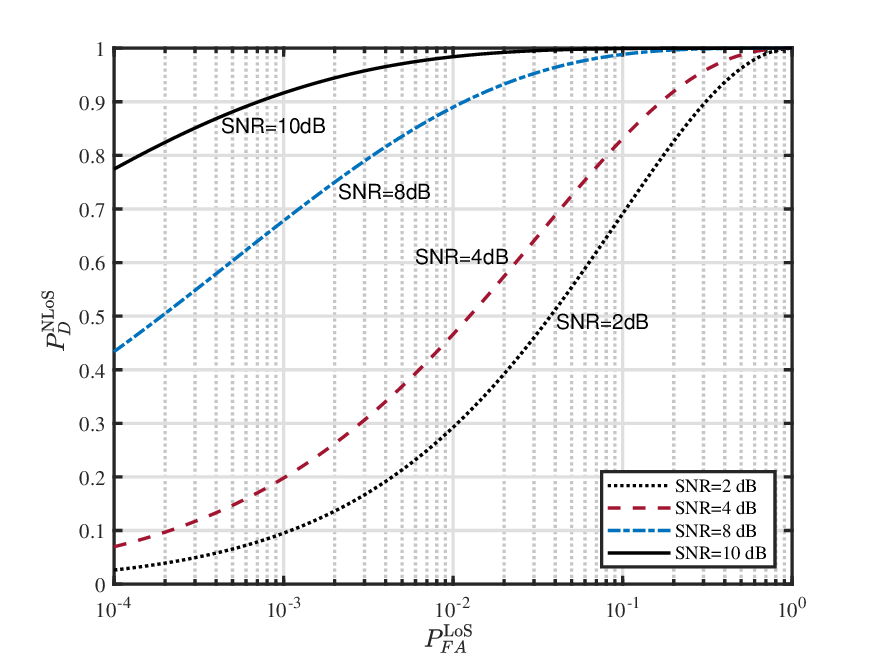}
\caption{Detection probability vs. false alarm probability given different SNRs.}
\end{minipage}
\end{figure}
\\
\indent
For simplicity, the simulated blockage events are randomly generated from a homogeneous Poisson process with constant $\lambda=10$ and $n=5$ without loss generality. For each experiment run, the distribution of the number of blocked channel realizations is independent of the time intervals, in which a fraction of trials are uniformly distributed. The results are shown in Fig. 3, where we generate 300 samples, where the strength of $\varepsilon$ is distinguished by different colors.
\\
\indent
According to the basic framework for binary hypothesis testing established in Section III-A, we further evaluate the performance of the proposed blockage detection method under different SNR conditions. Fig. 4 plots the detection probability versus false alarm probability given different SNRs, and demonstrates the effectiveness of the proposed blockage detection algorithm. It can be observed from Fig. 4 that with the SNR progressively increasing, the false alarm probability $P_{\text{FA}}^{\text{LoS}}$ is significantly reduced. For instance, assuming the optimal threshold to achieve a probability of detection $P_{\text{FA}}^{\text{LoS}}$ of 0.9, the probability of false alarm $P_{\text{FA}}^{\text{LoS}}$ drops from 0.1 (SNR=2 dB) to less 0.001 (SNR=10 dB). Moreover, we observe that a higher SNR can substantially improve the detection threshold, and SNR=10 dB is generally acceptable for blockage detection. In contrast, blockage detection becomes impossible in the low SNR regime due to the uncertainty caused by noise. These observations provide sufficient evidence that when the significance level of a specific alternative hypothesis is guaranteed, the proposed binary hypothesis test can efficiently reject the null hypothesis, and thereby is feasible to sense the random blockages while minimizing the false alarm risk.
\subsection{Outage Performance}
\begin{figure}[!t]
\centering
\begin{minipage}[t]{0.48\textwidth}
\centering
\includegraphics[width=8.5cm]{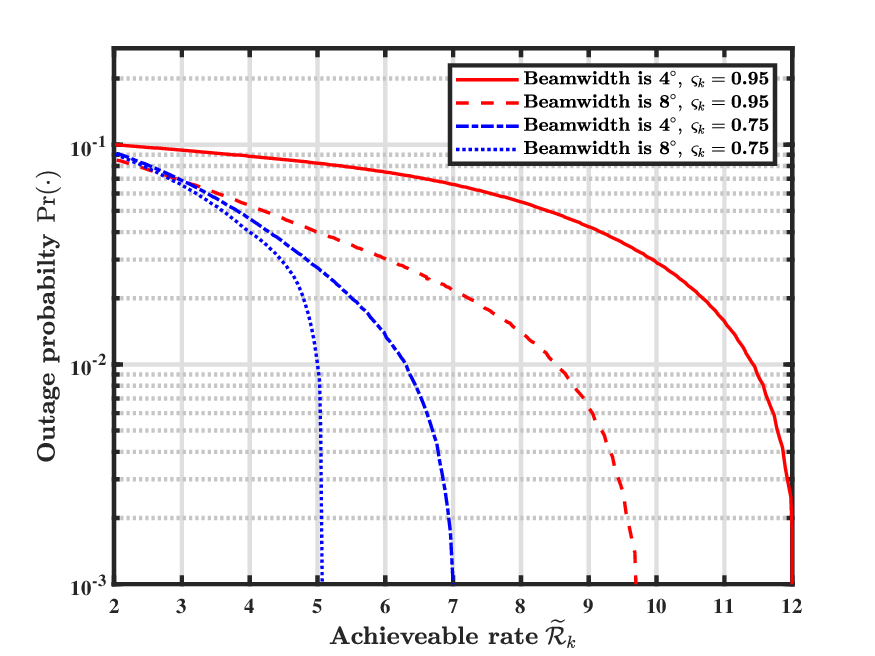}
\caption{Outage probability vs. average achievable rate, where $\mathcal N_k$ is set to be 128.}
\end{minipage}
\begin{minipage}[t]{0.48\textwidth}
\centering
\includegraphics[width=8.5cm]{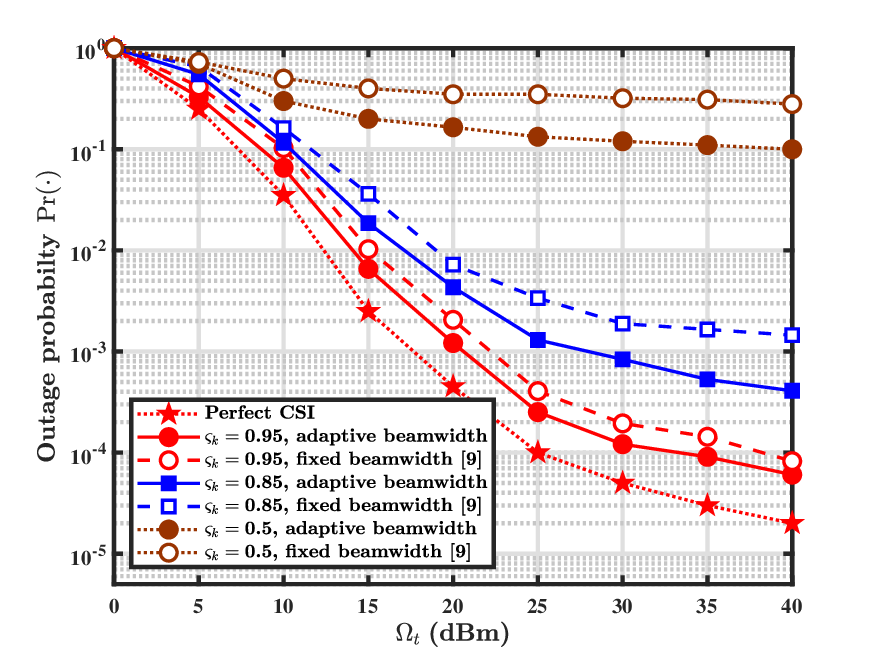}
\caption{Outage performance vs. $\Omega_t$.}
\end{minipage}
\end{figure}
\begin{figure}[!t]
\centering
\includegraphics[width=8.5cm]{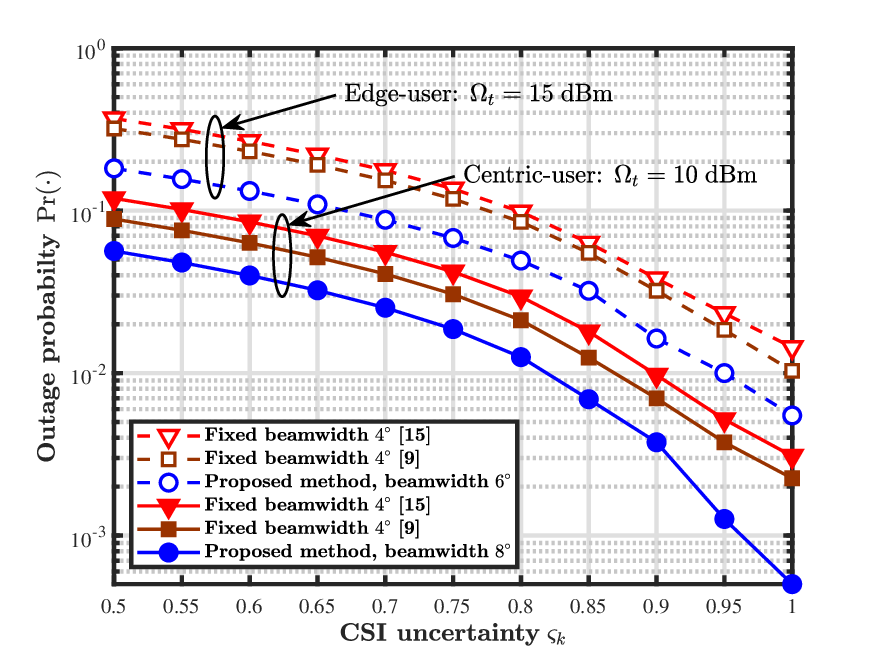}
\caption{Outage performance vs. CSI uncertainty, where different downlink power is allocated to the edge-user and the Centric-user.}
\label{figure}
\end{figure}
In Fig. 5, we further examine the role of the beamwidth in combating the CSI uncertainty and its effect on the average achievable rate of the downlink transmission. To visualize the relation between the beamwidth and CSI uncertainty, we plot outage probability $\Gamma_k$ versus achievable rate $\mathcal R_k$ and evaluate the outage performance of the proposed method under different levels of CSI uncertainty. It can be shown from Fig. 5 that the outage performance can be substantially improved by increasing the width of the reflecting beam. In fact, choosing a wider beam may be more advantageous, albeit at the expense of the achievable rate. In user mobility scenarios, when the LoS link is blocked, the wider reflecting beam is more beneficial for minimizing the outage probability, since it provides more stable and reliable connectivity. A wider beam implies that even though the estimates of the CSI and/or PSs are relatively coarse, passive beamforming at the RIS can retain highly reliable connectivity and achieve comparable performance. It is worth emphasizing that there is a trade-off between outage performance and achievable rate to overcome the limitation of imperfect CSI and allow for partial CSI. To be brief, if minimizing the outage probability is preferential and a bounded rate loss is acceptable, a wider beam is clearly a better choice, and vice versa.
\\
\indent
Next, Fig. 6 shows the outage probability as a function of the transmit power at the BS. We compare the outage performance of the proposed beamwidth control scheme with the fixed beamwidth solution given in \cite{Zhao21}. In our experiment, the reliability of estimate $s$ is set to be 0.95, 0.85, and 0.5, so as to evaluate the impact of the channel estimation error. From Fig.~6, one can observe that the outage performance of the proposed approach is better than that of the method in \cite{Zhao21} under various CSI uncertainty settings. With the CSI uncertainty merely exacerbating the channel estimation errors, it leads to the deterioration of the outage performance. Consequently, the beamwidth is critical in alleviating the impacts of the outage. The proposed adaptive beamwidth control method provides a trustworthy solution for tracking a mobile receiver. This is because a wider reflecting beam is capable of providing wider coverage, thereby providing resilient connectivity over the indirect RIS-MS link.
\\
\indent
In Fig. 7, we proceed to evaluate the outage probability versus CSI uncertainty in our proposed framework by taking into account the effects of the beamwidth and compare the performance against the methods in \cite{Zhao21} and \cite{Zhou21}.  In our experiment, we mainly consider two typical user access scenarios, namely the edge user and centric user. For a fair comparison, the distances between the edge- and centric-users to the RIS are respectively set to be 15 m and 30 m. From the presented results, we can see that the proposed technique can achieve better outage performance than those of the methods of \cite{Zhao21} and \cite{Zhou21}. It is interesting through scrutinizing this case that the centric-user with a wider and synthesizable beamwidth has a significant outage performance advantage over the edge-user. This observation suggests that even with inaccurate angle estimates, the proposed method can efficiently mitigate the impacts of the beam squint on the MS. It demonstrates the possibility that the proposed technique has the ability to provide a more resilient connection through RIS passive beamforming in the presence of blockages, which could be applied with imperfect CSI. It is important to realize that the wider and synthesizable beamwidth with a small amount of allocated power inevitably sacrifices the achievable rate during the transmitting procedure, and therefore an intrinsic trade-off needs to be considered in this case, examining such a trade-off is left for future work.
\subsection{Achievable Rate}
\begin{figure}[!t]
\centering
\begin{minipage}[t]{0.48\textwidth}
\centering
\includegraphics[width=8.5cm]{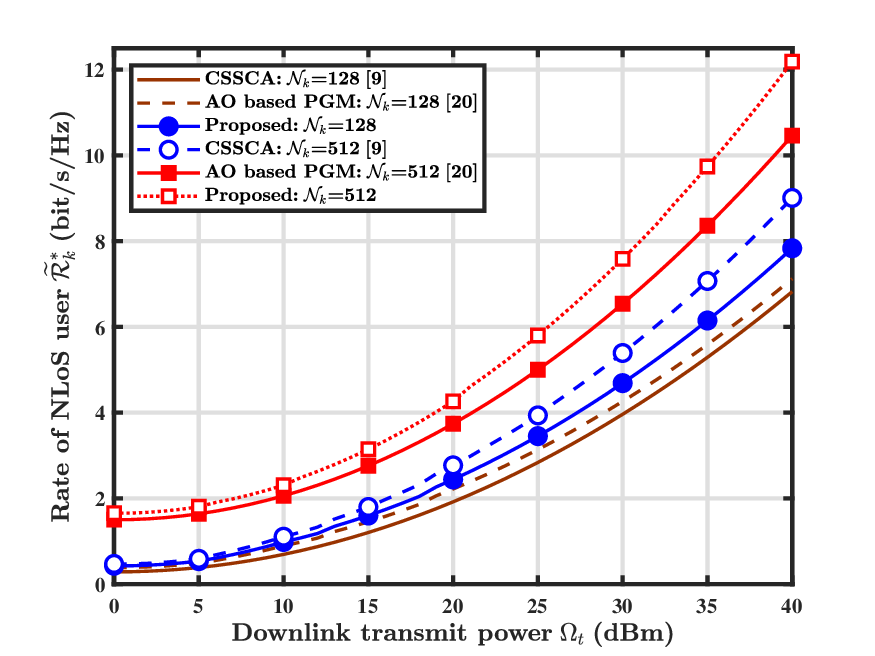}
\caption{Achievable rate of the accelerated PGD algorithm versus the CSSCA- and AO-based schemes.}
\end{minipage}
\begin{minipage}[t]{0.48\textwidth}
\centering
\includegraphics[width=8.5cm]{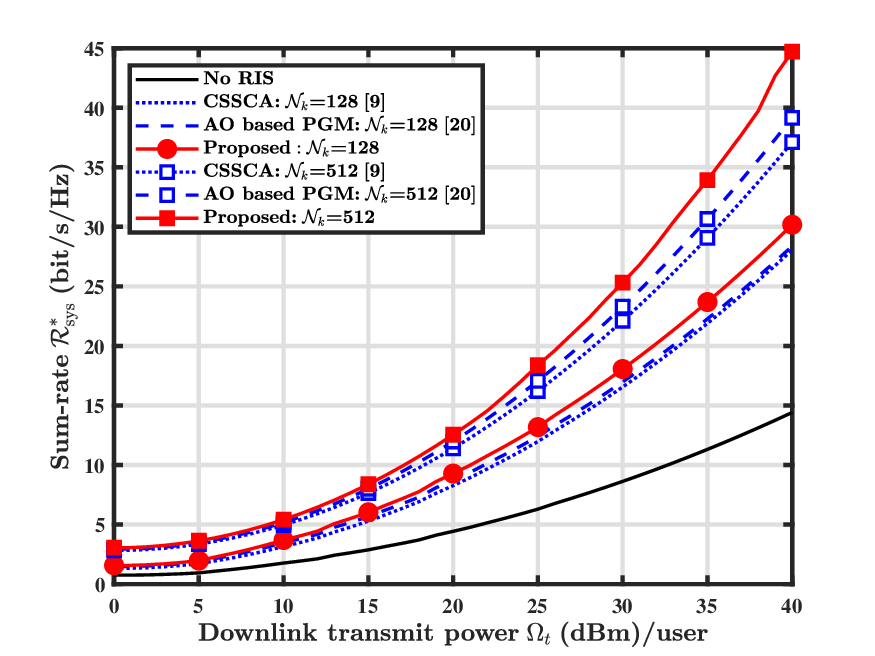}
\caption{Downlink sum-rate performance in an RIS-aided mmWave MIMO communication system, considering four active users and two of them being blocked.}
\end{minipage}
\end{figure}
\begin{figure}[!t]
\centering
\includegraphics[width=8.5cm]{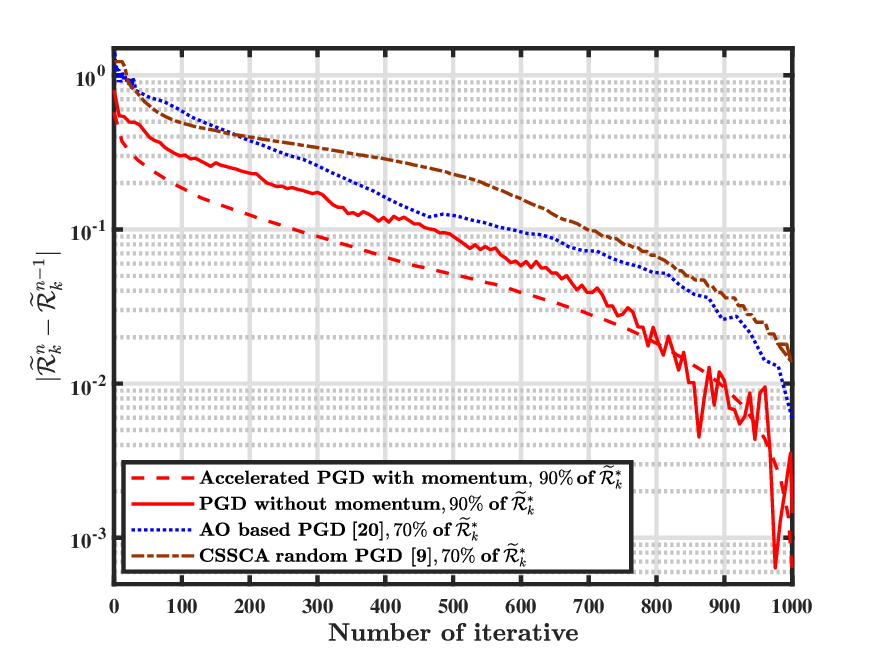}
\caption{Convergence rates of PGD yielded by different methods.}
\label{figure}
\end{figure}
For the following analysis, we further investigate the effectiveness of the proposed PGD-based optimization method for the RIS-aided transmission when the dominant LoS path is blocked. We analyze two cases for achievable rate ${\widetilde{\mathcal R}}_k$ and for downlink sum rate ${\mathcal R}_{\text{sys}}$. In the first case, the performance metric we consider is achievable rate $\mathcal R_k$, a function of the transmit power as formulated in (10). The performance is compared with references \cite{Zhao21} and \cite{Shi22}, with consideration of the impact of the number of RIS elements $\mathcal N_k$ on the achievable rate. Accordingly, the PGD optimization is performed in the scaled-variable form. Unless otherwise specified, we set parameters $\chi$=0.95 and the RIS-MS distance $D$=30 m, and therefore we can directly apply either the optimal transmission beam or the optimal reflecting beams toward an individual user. By examining the results presented in Fig. 8, we clearly see that the proposed PGD approach significantly boosts the achievable rate compared to the stochastic successive convex approximation (CSSCA) algorithm in \cite{Zhao21} and AO-based schemes in \cite{Shi22}. We also observe that scaling the number of RIS elements has a significant effect on the achievable rate. Intuitively, by increasing the number of PSs $\mathcal N_k$, the achievable rate increases. This is because the received signal power is proportional to the size of the RIS surface, such that not only the indirect link can be compensated by the increased number of PSs $\mathcal N_k$, but also the introduced RIS can create a narrower beam to provide a large directional gain.
\\
\indent
In the second case, we focus on evaluating the achievable sum rate in a multiuser scenario for the proposed PGD approach and the counterparts under the same consideration. Specifically, we consider a scenario where there are four active users, two of whom are completely blocked, whereby each user tries to receive data from the BS. For the sake of comparison, the communication mode of each user is still chosen on the basis of the settings in Fig. 8. We assume in further simulations that the transmission over the direct link is capable of providing the optimal achievable rate and maintains an approximately constant rate. We can see from Fig. 9 that without the aid of the RIS, or if the blockages cannot be effectively sensed, the sum rate of the system deteriorates severely. In contrast, we observe that with the aid of the RIS, all algorithms can achieve higher achievable sum rates, whereas the proposed PGD-based algorithm performs better than the CSSCA-based and AO-based algorithms. In addition, we study the impact of the number of PSs $\mathcal N_k$ on RIS-aided downlink transmission. As expected, the sum rate ${\mathcal R}^*_{\text{sys}}$ improves with increased $\mathcal N_k$ for each algorithm. It is evident that with a sufficiently large RIS, the proposed PGD-based algorithm can provide a near-optimal sum rate in the presence of blockages, similar to the case where the LoS link is available for all active users. 
\\
\indent
As a final numerical experiment, we investigate the convergence behavior of the proposed PGD-based algorithm compared with the aforementioned baseline algorithms. To further accelerate the convergence of the PGD-based algorithm, we incorporate the subspace constraint and the momentum term, as described in Sections IV-A and IV-B, respectively. The results are presented in Fig. 10, where the convergence rate of the objective gradient versus the number of iterations is plotted to depict the behavior of convergence to reach the optimal achievable rate. Our numerical results demonstrate that the proposed PGD-based algorithm with momentum acceleration outperforms other baseline algorithms, i.e., CSSCA-based and AO-based algorithms. It is evident from Fig. 10 that the subspace-constrained momentum-accelerated PGD algorithm is the fastest algorithm. For the case with only the subspace constraint, the proposed PGD-based method without momentum aid evidently achieves faster convergence and performs better than the AO-based strategies developed in \cite{Zhao21} and \cite{Shi22}, which run a gradient descent algorithm with a fixed step size to search the local optimum. After about 1,000 iterations, all algorithms get the stable minimum, corresponding to 70$\%$, 75$\%$, 90$\%$, and 90$\%$ of the desired optimal achievable rate $\widetilde{\mathcal R}_k^*$, respectively. Both momentum and subspace suffice convergence to the minimum in fewer iterations with higher achievable rates. Particularly, by introducing the momentum term, the perturbations are significantly reduced. This agrees with the intuition that reducing the momentum coefficients allows for faster convergence to be achieved. The reason for this phenomenon is that a constant step size is set to 0.05, while the convergence could readily get stuck due to a saddle point of the objective function. For conventional gradient descent methods, the larger step size enables faster convergence, but the risk of divergence is correspondingly higher. Experimental results illustrate that our proposed method can efficiently circumvent saddle points by taking an increasing momentum term, which is helpful in finding the global minimum. 
\section{Conclusion}
In this paper, we studied the NP hypothesis testing criterion and the momentum-based gradient optimization algorithm for solving the outage probability minimization and achievable rate maximization problems, respectively. Among them, the NP hypothesis testing provides a trustworthy means for random blockage detection. Our analysis highlighted the importance of the NP criterion-based detector in substantially alleviating the impact of blockages. Preliminary results also verified that with this integrated blockage sensing capability, the mmWave link availability becomes tractable. We further confirmed through numerical results that by leveraging the subspace-constrained and momentum-based optimization algorithms, the proposed PGD algorithm significantly outperforms the existing baseline algorithms and provably converges to the global optimum. The corresponding robust beamforming design can thus guarantee stable and reliable connectivity and has the possibility to unlock the full potential of mmWave communications. For future studies, we intend to take a multi-RIS setting into account to combat distributed blockages and to quantitatively investigate the intrinsic trade-off between the outage performance and the achievable rate. 
\ifCLASSOPTIONcaptionsoff
\newpage
\fi
%
\bibliography{IEEEabrv,RISPreference}
\end{document}